\documentclass[lettersize,journal]{IEEEtran}
\usepackage{amsmath,amsfonts,amssymb,amsthm}
\usepackage{array}
\ifCLASSOPTIONcompsoc
\usepackage[caption=false,font=normalsize,labelfon
t=sf,textfont=sf]{subfig}
\else
\usepackage[caption=false,font=footnotesize]{subfi
g}
\fi
\usepackage{textcomp}
\usepackage{stfloats}
\usepackage{url}
\usepackage{verbatim}
\usepackage{graphicx}
\usepackage{cite}
\usepackage{balance}
\usepackage{booktabs}
\usepackage{multirow}
\usepackage[ruled, linesnumbered]{algorithm2e}
\newtheorem{theorem}{Theorem}

\newtheorem{corollary}{Corollary}
\usepackage{pdfpages}
\usepackage{xcolor}
\usepackage{tabularx}
\usepackage{colortbl}
\usepackage{pdfpages}

\begin{document}

\title{Heterogeneous Interaction Modeling With Reduced Accumulated Error for Multi-Agent Trajectory Prediction}
\author{Siyuan Chen, and Jiahai Wang,~\IEEEmembership{Senior Member,~IEEE}
\thanks{
  This work is supported by the National Key R\&D Program of China (2018AAA0101203), the National Natural Science Foundation of China (62072483), and the Guangdong Basic and Applied Basic Research Foundation (2022A1515011690, 2021A1515012298). \textit{(Corresponding author: Jiahai Wang.)} 
  }
\thanks{
  Siyuan Chen and Jiahai Wang are with the School of Computer Science and Engineering, Sun Yat-sen University, Guangzhou 510275, China (e-mail: chensy47@mail2.sysu.edu.cn, wangjiah@mail.sysu.edu.cn).
  }
}

\markboth{IEEE Transactions on Neural Networks and Learning Systems}%
{Shell \MakeLowercase{\textit{et al.}}: A Sample Article Using IEEEtran.cls for IEEE Journals}


\maketitle

\begin{abstract}
  Dynamical complex systems composed of interactive heterogeneous agents are prevalent in the world, including urban traffic systems and social networks. Modeling the interactions among agents is the key to understanding and predicting the dynamics of the complex system, e.g., predicting the trajectories of traffic participants in the city. Compared with interaction modeling in homogeneous systems such as pedestrians in a crowded scene, heterogeneous interaction modeling is less explored. Worse still, the error accumulation problem becomes more severe since the interactions are more complex. To tackle the two problems, this paper proposes heterogeneous interaction modeling with reduced accumulated error for multi-agent trajectory prediction. Based on the historical trajectories, our method infers the dynamic interaction graphs among agents, featured by directed interacting relations and interacting effects. A heterogeneous attention mechanism is defined on the interaction graphs for aggregating the influence from heterogeneous neighbors to the target agent. To alleviate the error accumulation problem, this paper analyzes the error sources from the spatial and temporal perspectives, and proposes to introduce the graph entropy and the mixup training strategy for reducing the two types of errors respectively. Our method is examined on three real-world datasets containing heterogeneous agents, and the experimental results validate the superiority of our method.
\end{abstract}
\begin{IEEEkeywords}
Multi-agent trajectory prediction, heterogeneous interaction modeling, error accumulation, graph entropy, mixup training strategy.
\end{IEEEkeywords}

\section{Introduction}
\IEEEPARstart{M}{any} real-world complex systems, including social networks and urban traffic systems, can be regarded as dynamical systems composed of heterogeneous interacting agents. Different types of agents present various intrinsic behavior patterns, and the dynamic interactions among agents are even more complex, leading to complicated dynamics at both the individual level and the system level. Understanding the interactions among heterogeneous agents can help us predict and control the behavior of a system~\cite{hu2021distributed,zhang2017optimal,du2021multiagent}. A typical application is trajectory prediction, which is a fundamental task in autonomous driving and mobile robot navigation. 
\begin{figure}[t]
  \centering
  \includegraphics[width=0.98\columnwidth]{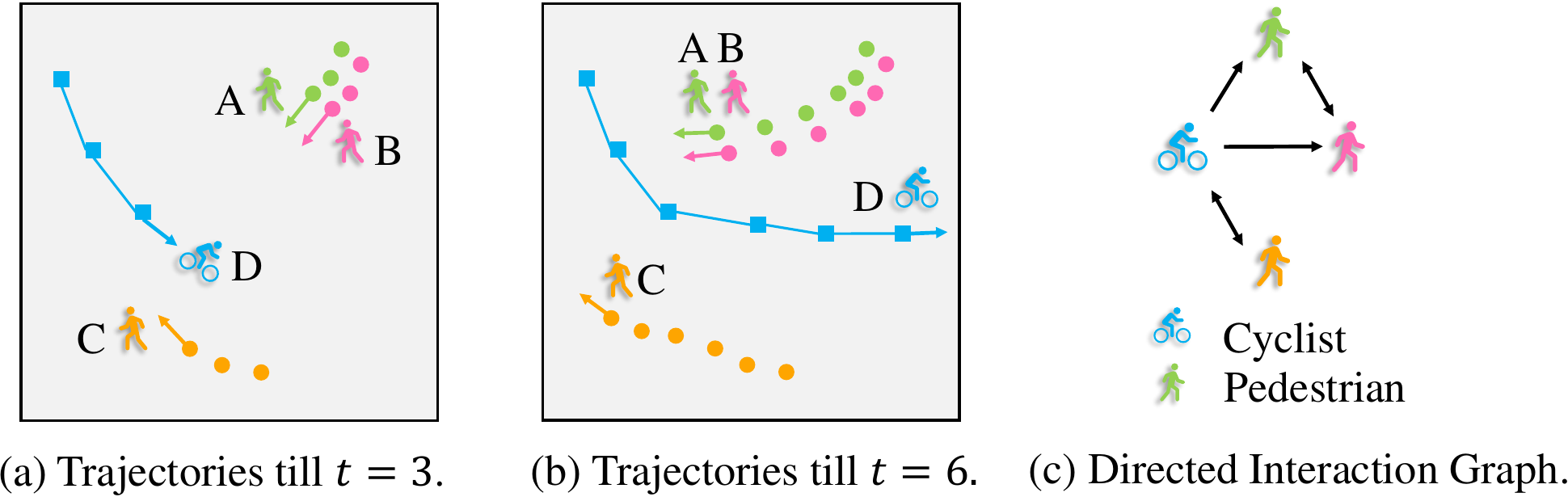}
  \caption{A heterogeneous dynamical system involving three pedestrians (A, B and C) and one cyclist (D). Two snapshots of the trajectories are taken at time $t=3$ (a) and $t=6$ (b). A possible directed interaction graph among all traffic participants is provided (c).}
  \label{fig:example}
\end{figure}

Compared with \textit{homogeneous} systems, the main difficulty of \textit{heterogeneous} multi-agent trajectory prediction lies in the heterogeneity of agents$'$ interactions~\cite{ivanovic2019trajectron,li2020evolvegraph}. Inaccurate interaction modeling at each time step, in turn, leads to a more severe error accumulation problem in multi-step prediction.

The first problem, heterogeneity of agents$'$ interactions, is that the interacting patterns are diversified among heterogeneous agents. This phenomenon is less explored in pedestrian trajectory prediction~\cite{kothari2021human}, where researchers focus on \textit{homogeneous} interaction modeling. Usually, researchers assumed that distance is a consistent indicator of interactions. They built an interaction graph based on the pairwise distances, and then apply social pooling \cite{alahi2016social}, attention mechanisms \cite{yu2020spatio} or graph neural networks (GNN) \cite{mohamed2020social} to capture the spatial dependence among agents induced by the interaction graph. However, the distance may not be an accurate indicator of interactions \cite{shi2021sgcn,zhou2021ast}, especially in \textit{heterogeneous} systems. A motivating example involving three pedestrians and one cyclist is provided in Fig.~\ref{fig:example} to illustrate this problem. 

This paper gains three observations from Fig.~\ref{fig:example} as follows. \textbf{(I)} At time $t=3$, Pedestrian A and B walk in a group, and the distance between them suggests a strong interaction. Pedestrian C is distant from A and B, indicating fairly weak interactions. \textbf{(II)} Although the distance between Cyclist D and Pedestrian C is larger than that between Pedestrian A and B, their interaction is strong as they tend to avoid a collision. \textbf{(III)} Pedestrian A and B change their walking directions in advance as they try to avoid meeting Cyclist D. By contrast, Cyclist D is less affected by the two pedestrians. To sum up, the distance may not be a consistent indicator of interactions in \textit{heterogeneous} systems. Besides, as a symmetric metric, distance can fail to capture asymmetric interactions.

Therefore, for \textit{heterogeneous} interaction modeling, it is more preferable to learn a latent interaction graph~\cite{li2020evolvegraph} other than using a pre-defined graph. Furthermore, the information aggregation over the graphs should be able to distinguish interacting effects between heterogeneous agents. Li \textit{et al.}~\cite{li2020evolvegraph} proposed to learn multi-relational graphs, but the number of relation types is hard to decide in practice. Ivanovic \textit{et al.}~\cite{ivanovic2019trajectron} defined an edge type as the concatenation of node types, but the corresponding category-aware modules suffered from quadratic space complexity. Thus, it remains to explore a more appropriate construction of interaction graphs and a lightweight design for heterogeneous information aggregation over the graphs.

The second problem, error accumulation, is an inevitable issue even in homogeneous systems. A wide class of models makes multi-step predictions recursively, i.e., predicting an agent$'$s next position based on its previous predictions. In these models, errors are aggregated from neighboring agents and accumulate over multiple time steps, leading to a possible large deviation in the long-term prediction. Nevertheless, to the best of our knowledge, few efforts have been paid to this issue in the field of trajectory prediction. 

Error accumulation can be effectively eliminated in the training stage by feeding the true positions of agents to the model, i.e., teacher forcing~\cite{williams1989learning}, but this will give rise to a large discrepancy between testing and training. Such phenomenon is well-known as the exposure bias for auto-regressive language models in natural language generation \cite{ranzato2015sequence}. Zhang \textit{et al.} \cite{zhang2019bridging} proposed to select oracle words that simulate the ground truth words, and randomly switched between the oracle words and the ground truth during the training procedure. Their method proved to be effective in narrowing the gap between training and inference. This inspires us to mix up the true positions and some high-quality candidates as the input of models for reducing the accumulated error.

To deal with the heterogeneity of interactions and the error accumulation issue, this paper proposes \textbf{H}eterogeneous \textbf{I}nteraction \textbf{M}odeling with \textbf{R}educed \textbf{A}ccumulated \textbf{E}rror (HIMRAE) for multi-agent trajectory prediction. Our method adopts an encoder-decoder framework. The encoder dynamically infers two types of information from the historical trajectories, the directed interacting relations that tell if one agent affects another, and the interacting effects that implicitly describe how the interaction works. The interacting relations and the interacting effects jointly define an edge-featured interaction graph. The decoder defines a heterogeneous attention mechanism over the interaction graph to model the interactions among heterogeneous agents, and finally, it predicts future trajectories using category-aware recurrent neural networks. 

Regarding the recursive multi-step prediction in multi-agent systems as the forward pass of a multi-layer graph neural network, this paper analyzes the error accumulation problem from the spatial and temporal aspects. To reduce the spatially aggregated errors, the graph entropy is introduced for penalizing the complexity of the interaction graphs. The minimizer of graph entropy is characterized in theory to guide the choice of regularization coefficient. For the temporally accumulated errors, this paper proposes to mix up the ground truth trajectories and the predicted ones in the training stage, which balances the model$'$s ability in accurate single-step prediction and multi-step prediction. The mixup training strategy is proved to enjoy a lower error bound in theory.

The contributions of this paper are summarized as follows. 

\begin{itemize}
  \item An encoder-decoder framework is proposed for heterogeneous multi-agent trajectory prediction. Our method dynamically infers the interacting relations and the interacting effects among agents, characterizing the interaction structure as an edge-featured graph.
  \item This paper proposes a heterogeneous attention mechanism of linear space complexity over the interaction graphs to model the complex heterogeneous interactions.
  \item This paper analyzes the error accumulation problem in multi-agent trajectory prediction from both the spatial and temporal perspectives, and proposes to use the graph entropy and a mixup training strategy to reduce the two types of errors respectively. Both the graph entropy and the mixup training strategy have theoretical justifications under some simplified assumptions.
\end{itemize}

\section{Related Works}
\subsection{Interaction Modeling in Multi-Agent Trajectory Prediction}
Traffic participants, like pedestrians and vehicles, interact with others in the surroundings and adjust their trajectories towards the destinations. The interaction is complex itself, and evolves over time, leading to the complicated trajectories of multiple traffic participants. Interacting modeling for multi-agent systems generally includes interaction graph construction and information aggregation over the graph. The techniques differ in \textit{homogeneous} and \textit{heterogeneous} systems.

In \textit{homogeneous} systems, grids~\cite{alahi2016social}, complete graphs~\cite{huang2019stgat,yang2021graphbased}, distance based graphs~\cite{mohamed2020social,yu2020spatio} were used as pre-defined interaction graphs. Since complete graphs and distance-based graphs contain superfluous edges and fail to model directed effects, some works~\cite{kipf2018neural,shi2021sgcn,zhou2021ast} proposed to learn asymmetric latent graphs. Given the interaction graphs, social pooling~\cite{alahi2016social}, spatio-temporal graph convolutions, spatio-temporal graph attention~\cite{huang2019stgat,yu2020spatio} were introduced to aggregate the interacting effects. Prior knowledge like sparsity~\cite{shi2021sgcn} and obstacle avoidance~\cite{yang2021graphbased} were used for attention computation.

In the \textit{heterogeneous} systems, the interaction graphs can be extended to heterogeneous graphs with different types of nodes and edges. The edge types can be pre-defined as the concatenation of node types~\cite{ivanovic2019trajectron,salzmann2020trajectron++}, e.g., ``Pedestrian-Bus'', or learned by the model from observational data~\cite{li2020evolvegraph}. Relative position, velocity, and heading angle can be further incorporated as edge features~\cite{li2021spatio,mo2021heterogeneous}.

Given the heterogeneous interaction graphs, node type aware modules were used for trajectory encoding, type-specific supernodes were used to share information among agents of the same type~\cite{ma2019trafficpredict,li2021hierarchical}, and edge type aware modules were used for modeling interacting effects, which are aggregated to predict future states~\cite{ivanovic2019trajectron,salzmann2020trajectron++}. However, there are some drawbacks of edge type aware aggregations. When the edge type is defined by concatenating node types~\cite{ivanovic2019trajectron,salzmann2020trajectron++}, the number of edge type aware modules grows quadratically, which can be a redundant design. When the edge type is learned by the model~\cite{li2020evolvegraph}, the number of edge types is hard to decide in practice. Besides, some works~\cite{ma2019trafficpredict, salzmann2020trajectron++} limited themselves to a pre-decided set of agent types when coding the category-aware modules, which is less flexible when agent types are different in new datasets.

Unlike existing works, this paper proposes to learn edge-featured graphs, where an edge indicates a directed interaction, and an edge feature describes the interacting effects without defining edge types. Besides, a heterogeneous attention mechanism of linear space complexity is proposed to aggregate the interacting effects. 

\subsection{Heterogeneous Graph Neural Networks}

Graphs with more than one type of nodes or edges, termed as heterogeneous graphs or heterogeneous information networks~\cite{shi2016survey,ji2021heterogeneous}, widely exist in the world, such as social networks, knowledge graphs and proteins. GNNs developed for this form of data are called heterogeneous graph neural networks. The message passing mechanisms of homogeneous GNNs~\cite{gilmer2017neural} were substantially extended by introducing category-aware modules for different types of nodes~\cite{zhang2019heterogeneous}, edges~\cite{schlichtkrull2018modeling} and their compositions, meta-paths~\cite{wang2019heterogeneous}. Representative works include relation type aware aggregators~\cite{schlichtkrull2018modeling}, heterogeneous graph transformer~\cite{hu2020heterogeneous}, hierarchical aggregators defined over node types~\cite{zhang2019heterogeneous} and hierarchical attention defined over both node-level and semantic-level~\cite{wang2019heterogeneous}.

Despite the great success of heterogeneous graph neural networks in heterogeneous graphs, these methods are not applicable to heterogeneous multi-agent trajectory prediction. The key difference is that there are no natural well-defined relations between two agents. Even when the categories of the agents and the relative position between them are known, the interactions can be variable. Besides, these methods are not designed to integrate possible edge features. Our method can infer the relations between agents and model their interactions over an edge-featured graph.

\subsection{Learning Graphs from Data}
Graphs provide a structured representation of data. When the graph representation is not readily available, researchers seeked to learn it from observational data~\cite{dong2019learning}, with application to multi-variate time series forecasting~\cite{wu2020connecting} and reasoning over physical systems~\cite{kipf2018neural}. Edge directions~\cite{wu2020connecting,kipf2018neural}, edge types~\cite{kipf2018neural}, and co-existence of edges~\cite{chen2021neural} were considered for graph learning. The learning is often guided by a task-specific objective, while the complexity of the graph is not explicitly controlled. Sparsity~\cite{kipf2018neural} is an intuitive choice to penalize the graph complexity. However, a simpler graph is not necessarily sparser, and a sparser graph may not necessarily lead to better performance. Instead, this paper introduces another metric, the graph entropy, to reduce the graph complexity, which helps reduce the prediction error.

\section{Method: HIMRAE}
\begin{figure*}[t]
  \centering
  \includegraphics[width=0.95\textwidth]{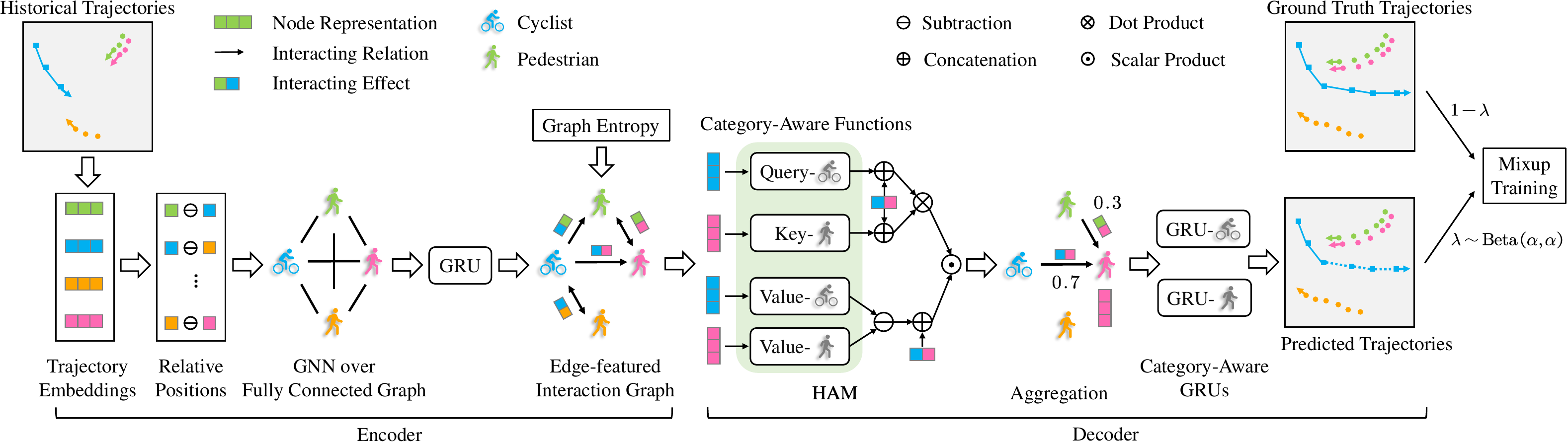}
  \caption{Overview of HIMRAE. HIMRAE takes the form of an auto-encoder. The encoder applies a GNN with a GRU over the historical trajectories to infer dynamic interaction graphs. The interaction graph is an edge-featured graph, where an edge represents a directed interacting relation and an edge feature represents the interacting effect. The graph entropy is used to control the structural complexity of the interaction graph. The decoder defines a heterogeneous attention mechanism (HAM) to model the interactions among heterogeneous agents, and predicts future trajectories recursively. The predicted trajectories and the ground truth trajectories are mixed with a coefficient $\lambda$ sampled from a beta distribution $\text{Beta}(\alpha,\alpha)$ to reduce the temporally accumulated error.}
  \label{fig:overview}
\end{figure*}
\subsection{Problem Definition}
In an $N$-agent heterogeneous dynamical system, each agent $i$ belongs to a category $c_i$ out of $C$ categories. The number of agents may vary case by case. The 2D coordinate of agent $i$ at time $t$ is recorded as $\mathbf{x}^t_i=(x^t_i,y^t_i)$. Let $T_h$ and $T_f$ denote the number of historical steps and future steps, respectively. The problem of trajectory prediction is that, given the historical trajectories of all agents $\mathbf{X}^{1:T_h}=\{\mathbf{x}^{1:T_h}_i\}^N_{i=1}$, predict their future trajectories $\widehat{\mathbf{X}}^{T_h+1:T_h+T_f}$ that best match the ground truth trajectories $\mathbf{X}^{T_h+1:T_h+T_f}$. A summary of the notations used in this paper can be found in \textit{Appendix~A}.
\subsection{Hypotheses, Motivation and Overview of HIMRAE}
The trajectory of each agent is affected by the agents it interacts with, and the interactions vary over time. The interactions among agents at time $t$ can be abstracted as a directed graph $\mathcal{G}^t=(\mathcal{V},\mathcal{E}^t)$, with each agent $i\in\mathcal{V}$ as a node and the interacting relation from agent $i$ to agent $j$ as an directed edge $(i,j)\in\mathcal{E}^t$. Generally, the underlying interaction graphs are unobservable, but they can be inferred from the historical trajectories and used for future trajectory prediction.

This paper hypothesizes that the interacting relations among agents within a small time window are relatively stable. Therefore, a $\theta$-parameterized distribution of the full trajectories can be factorized as
\begin{equation}
  \begin{aligned}
    p_\theta\left(\mathbf{X}^{1:T_h+T_f}\right)
    &= p_\theta\left(\mathbf{X}^{1:\tau}\right)\prod^M_{m=1}p_\theta\left(\mathbf{X}^{1+m\tau:(m+1)\tau}|\mathbf{X}^{1:m\tau}\right)\\
    &\quad \times p_\theta\left(\mathbf{X}^{1+M\tau:T_h+T_f}|\mathbf{X}^{1:M\tau}\right),
  \end{aligned}  
\end{equation}
where $\tau$ is the window size, and $M=\lfloor (T_h+T_f)/\tau \rfloor$ is the maximum number of the time windows. An appropriate $\tau$ may be related to the number of agents $N$, since the interactions change more frequently when $N$ is large. Nonetheless, $\tau$ is assumed to be case agnostic in this paper for simplicity, and a more sophisticated design is left for future work.

To motivate the design of the interaction graph, the authors observe that: in practice, one has little prior knowledge about the semantics of the interaction between two agents, e.g., the type of the interaction and its effect, even when the agent types and their relative position are known. Therefore, this paper seeks to infer the existence of interactions and a hidden state that implicitly encodes the interacting effects. Formally, given the trajectories in the previous time window, one can infer the interacting relations and interacting effects among agents, which lead to the following factorization, 
\begin{equation}
  \begin{aligned}
    &\quad\,\, p_\theta\left(\mathbf{X}^{1+m\tau:(m+1)\tau}|\mathbf{X}^{1:m\tau}\right)\\
    &= \int p_\theta\left(\mathbf{X}^{1+m\tau:(m+1)\tau}|\mathbf{X}^{1:m\tau},\mathbf{Z}^m,\mathbf{E}^m\right)\\
    &\quad\times p_\theta\left(\mathbf{Z}^m,\mathbf{E}^m|\mathbf{X}^{1:m\tau}\right)\,\text{d}\mathbf{Z}^m\,\text{d}\mathbf{E}^m,
  \end{aligned}  
\end{equation}
where $\mathbf{Z}^m$ and $\mathbf{E}^m$ are the interacting relations and interacting effects in the $m$-th time window, respectively. $\mathbf{Z}^m\in\{0,1\}^{N\times N}$ is a zero-diagonal binary matrix with $z^m_{ij}=1$ indicating an directed edge $(i,j)$. $\mathbf{E}^m\in\mathbb{R}^{N\times N\times D}$ is a real-valued tensor with $\mathbf{e}^m_{ij}$ representing the impact of agent $i$ on agent $j$. Regarding $\mathbf{E}^m$ as the edge features, $\mathbf{Z}^m$ and $\mathbf{E}^m$ jointly define an edge-featured interaction graph.

Directly evaluating $p_\theta\left(\mathbf{Z}^m,\mathbf{E}^m|\mathbf{X}^{1:m\tau}\right)$ is intractable~\cite{VAE}, therefore, a neural network $q_\phi\left(\mathbf{Z}^m,\mathbf{E}^m|\mathbf{X}^{1:m\tau}\right)$ parameterized by $\phi$ is used as an approximation. Interpreting $q_\phi$ and $p_\theta$ as an encoder and a decoder respectively, our method falls in the encoder-decoder framework.

The predicted trajectories $\widehat{\mathbf{X}}$ are assumed to obey an isotropic Gaussian distribution. Therefore, maximizing the log-likelihood conditioned on the historical trajectories leads to minimizing the expected squared loss,
\begin{equation}
  \begin{aligned}
    \mathcal{L}
    &=\underset{\mathbf{Z}^{1:M},\mathbf{E}^{1:M}|\mathbf{X}^{1:T_h}}{\mathbb{E}}\left[\frac{1}{NT_f}\sum^{T_h+T_f}_{t=T_h+1}\left\|\mathbf{X}^t-\widehat{\mathbf{X}}^t\right\|^2_F\right].
  \end{aligned} 
  \label{eq:reconstruction_error}
\end{equation}

Though intuitive, the formulation above raises two issues: (I) Unlike the variational auto-encoder (VAE)~\cite{VAE}, it lacks a natural choice for constraining the latent variables, as there is little prior knowledge about the interaction graph in practice. (II) Directly optimizing $\mathcal{L}$ will face a severe problem of error accumulation from the recursive nature of the decoder. Since a simplified interaction graph can be more controllable, this paper adopts the graph entropy to penalize the structural complexity of the graph, which also alleviates the error propagation problem in the spatial domain. For the temporally accumulated errors, this paper introduces a mixup training strategy.

An overview of the proposed method is shown in Fig.~\ref{fig:overview}. The details of the encoder and decoder are described in Section~\ref{sec:encoder} and Section~\ref{sec:decoder}, respectively. Section~\ref{sec:graph_entropy} analyzes the error sources from the spatial and temporal domains, and describes the usage of the graph entropy. Section~\ref{sec:mixup} describes the mixup training strategy. Section~\ref{sec:charateristics} summarizes the charateristics of our method.
\subsection{Encoder}\label{sec:encoder}
The encoder aims at inferring the dynamically evolving interaction graph in each time window. Since the underlying graph is unknown, a GNN is applied over a fully connected graph to learn the latent interaction structure, and a gated recurrent unit (GRU)~\cite{GRU} is used to capture the temporal dependence of the latent graphs. The structure of the encoder is visualized on the left of Fig.~\ref{fig:overview}.

Firstly, the historical trajectories are embedded to a latent space to obtain the node representation of each agent,
\begin{equation}
  \mathbf{v}^m_j = f_{\text{emb}}(\mathbf{x}_j^{1+(m-1)\tau:m\tau}),
\end{equation}
where $f_{\text{emb}}$ is a multi-layer perceptron (MLP). Then, one can use the relative position $\mathbf{v}^m_i - \mathbf{v}^m_j$ between two agents in the latent space to describe their spatial relationship. $\mathbf{v}^m_i - \mathbf{v}^m_j$ measures the correlation between the trajectories of two agents, which can be a good indicator of their interaction. This is advantageous over the stepwise relative position in the original space\cite{mohamed2020social}, a local metric that may fail to capture the global correlation.

The relative position can be viewed as a message over an edge. Following the message passing formulation of GNNs~\cite{gilmer2017neural}, a two-layer GNN is designed as follows,
\begin{align}
  \widetilde{\mathbf{v}}^m_j &= {\textstyle f_v\left(\sum_{i\neq j}f_e(\mathbf{v}^m_i-\mathbf{v}^m_j)\right),}\\
  \widetilde{\mathbf{e}}^m_{ij} &= \widetilde{f}_e(\widetilde{\mathbf{v}}^m_i-\widetilde{\mathbf{v}}^m_j), 
\end{align}
where $f_v$ is an MLP updating the node embeddings, $f_e$ and $\widetilde{f}_e$ are MLPs that updates the edge embeddings. Considering the uncertainty of the interacting effect, $\mathbf{e}^m_{ij}$ is sampled from a Gaussian distribution $\mathcal{N}(\widetilde{\mathbf{e}}^m_{ij},\mathbf{I})$.

As for the interacting relation, a GRU is used to model its evolution over time, i.e.,
\begin{equation}
  \mathbf{r}^m_{ij} = \text{GRU}(\widetilde{\mathbf{e}}^m_{ij},\mathbf{r}^{m-1}_{ij}), 
\end{equation}
where $\mathbf{r}^{m}_{ij}$ is an edge representation encoding the dynamics of the interacting relation. By projecting $\mathbf{r}^m_{ij}$ to a scalar, one can calculate the interacting probability from one agent to another. However, sampling the interacting relation is non-differentiable since $z^m_{ij}$ is a discrete variable. Fortunately, the Bernoulli distribution has a continuous approximation named the binary concrete distribution~\cite{Gumbel-softmax} that allows back-propagation. Formally, $z^m_{ij}$ is sampled via the following reparameterization trick,
\begin{equation}
  z^m_{ij} = \text{Sigmoid}((f_{\text{proj}}(\mathbf{r}^m_{ij})+\ln \delta - \ln (1-\delta))/T),\label{eq:gumbel}
\end{equation}
where $\delta\in\mathbb{R}$ is drawn from the $\text{Gumbel}(0, 1)$ distribution and $T$ is a temperature parameter that controls the ``smoothness'' of the samples. $f_{\text{proj}}$ maps $\mathbf{r}^m_{ij}$ to a scalar.

\subsection{Decoder}\label{sec:decoder}
The decoder is intended for modeling the interactions among heterogeneous agents based on the interaction graphs, and predicting their future trajectories. A heterogeneous attention mechanism (HAM) is used to capture spatial dependence among heterogeneous agents, and category-aware GRUs are used to capture agents$'$ intrinsic dynamics. The procedure of the decoder is visualized on the right of Fig.~\ref{fig:overview}.

To distinguish the importance of different types of agents, this paper extends the scaled dot-product attention in Transformer~\cite{vaswani2017attention} by introducing category-aware modules. The attention mechanism is defined as mapping a query and a set of key-value pairs to an output. Given an edge $(i,j)$, this paper treats the source node $i$ as a ``query'' and the target node $j$ as a ``key'' to compute the attention score. The relative position concatenated with the interacting effect is regarded as the ``value'' corresponding to the key. The sum of the values from adjacent agents weighted by the attention scores represents the overall influence of interactions. To model the effect of heterogeneous agents, this paper simply applies category-aware mappings to the agent representations before the aforementioned computation. 

Let $\mathbf{h}^t_j$ be the hidden vector encoding the dynamics of agent $j$ at time $t$ in the decoder. A heterogeneous attention mechanism is formulated as follows,
\begin{equation}
  \mathbf{m}_j^t = {\textstyle \sum_{i\neq j}\alpha^t_{ij}\cdot f_{V}\left(\left[g^{c_i}_{V}(\mathbf{h}^t_i)-g^{c_j}_{V}(\mathbf{h}^t_j),\mathbf{e}^m_{ij}\right]\right),}\label{eq:value}
\end{equation}
where $\mathbf{m}_j^t$ is the aggregated interacting effects from agent $j'$ s neighbors, and $[\cdot,\cdot]$ is the concatenation operator. $f_{V}$ is an MLP updating the value vector, and $g^{c_i}_{V}$ is a category-aware single-layer perceptron that maps different types of agents to a common space. $\alpha^t_{ij}$ is the attention score defined in a softmax-like form,
\begin{equation}
  \alpha^t_{ij} = \frac{z^m_{ij}\cdot \exp (a^t_{ij})}{\sum_{z^m_{ij} > 1/2}\,\, z^m_{ij}\cdot \exp (a^t_{ij})}.
  \label{eq:att_normalization}
\end{equation}
In the training stage, $z^m_{ij}$ sampled via Eq.~(\ref{eq:gumbel}) ranges in $[0, 1]$, and only edges with $z^m_{ij}>1/2$ are involved in the computation. In the testing stage, Eq.~(\ref{eq:att_normalization}) is the normal graph attention over the inferred edges without ambiguity.

In Eq.~(\ref{eq:att_normalization}), $a^t_{ij}$ is calculated via the scaled inner product,
\begin{equation}
  a^t_{ij}= \frac{1}{\sqrt{D}}f_{Q}\left(\left[g^{c_i}_{Q}(\mathbf{h}^t_i),\mathbf{e}^m_{ij}\right]\right)^Tf_{K}\left(\left[g^{c_j}_{K}(\mathbf{h}^t_j),\mathbf{e}^m_{ij}\right]\right).\label{eq:query_key}
\end{equation}
$f_{Q}$ and $f_{K}$ are single-layer perceptrons for updating the query vector and the key vector, respectively. $g^{c_i}_{Q}$ and $g^{c_i}_{K}$ are category-aware mappings like $g^{c_i}_{V}$. $D$ is the dimension of the query vector. In Eqs. (\ref{eq:value})-(\ref{eq:query_key}), $\mathbf{e}^m_{ij}$ implicitly decides the interacting effect without explicitly predefining a finite set of interacting relations. HAM can also be extended to incorporate sub-categories for modeling finer-grained or personalized trajectory patterns. The sub-categories can be learned by clustering~\cite{xing2020personalized} or contrastive learning~\cite{chen2021personalized} that encourages discriminative representations, which are left for future works.

Note that HAM is a natural extension of the self-attention in Transformer to an edge-featured graph containing multiple types of nodes. By dropping the category-aware mappings $g^{c_i}_{Q/K/V}$, HAM reduces to a homogeneous attention mechanism. By further ignoring the edge feature $\mathbf{e}^m_{ij}$, the original self-attention is recovered. Besides, HAM is a lightweight design for heterogeneous interaction modeling since the number of category-aware modules is linear of the agent types.

After modeling the spatial dependence among agents in HAM, our method use category-aware GRUs~\cite{li2020evolvegraph} to capture the intrinsic behavior patterns for different types of agents,
\begin{equation}
  \mathbf{h}^{t+1}_j = \text{GRU}_{c_j}\left(\left[\mathbf{m}_j^t,\widehat{\mathbf{x}}^{t}_j\right], \mathbf{h}^{t}_j\right).
  \label{eq:cat_gru}
\end{equation}
The hidden vector $\mathbf{h}^{t+1}_j$ is used to predict future trajectories,
\begin{equation}
  \boldsymbol\mu^{t+1}_j = \widehat{\mathbf{x}}^{t}_j + f_{\text{out}}(\mathbf{h}^{t+1}_j+\boldsymbol\epsilon),
  \label{eq:mu}
\end{equation}
where $f_{\text{out}}$ is an MLP outputting the change in positions, and $\boldsymbol\epsilon\sim\mathcal{N}(\boldsymbol 0,\mathbf{I})$ is a noise vector to increase diversity. $\widehat{\mathbf{x}}^{t+1}_j$ is assumed to follow an isotropic Gaussian distribution $\mathcal{N}(\boldsymbol\mu^{t+1}_j,\sigma^2\mathbf{I})$ with $\sigma^2$ as a fixed variance. The mean $\boldsymbol\mu^{t+1}_j$ is used as the predicted positions. As is done in EvolveGraph~\cite{li2020evolvegraph}, the predicted position $\widehat{\mathbf{x}}^{t}_j$ in Eqs.~(\ref{eq:cat_gru})-(\ref{eq:mu}) is replaced by the ground truth $\mathbf{x}^{t}_j$ during the historical steps.

A recent work, reinforced hybrid attention inference network (RAIN)~\cite{li2021rain}, shares similar a pipeline on latent graph learning with our method. RAIN first uses reinforcement learning to identify important relations that decrease the prediction error, and then assigns the relations different attention weights. Unlike RAIN, our method uses the reparameterization trick to allow back-propagation, offering a simpler option for end-to-end training without a specific design for reinforcement learning. Besides, the soft attention mechanism in RAIN is still homogeneous, while ours is heterogeneous.

\subsection{Loss Function with Graph Entropy Regularization}\label{sec:graph_entropy}
\begin{figure*}[!t]
  \centering
  \subfloat[Error accumulation problem.]{\includegraphics[width=0.245\textwidth]{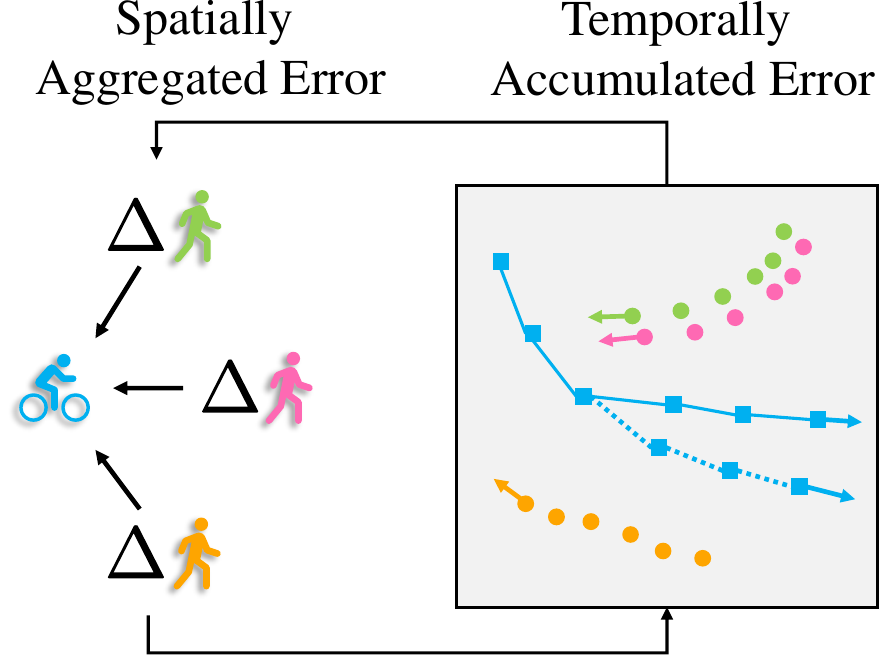}}
  \label{fig:error_accumulation}
  \hfil
  \subfloat[Graph entropy for regularization.]{\includegraphics[width=0.23\textwidth]{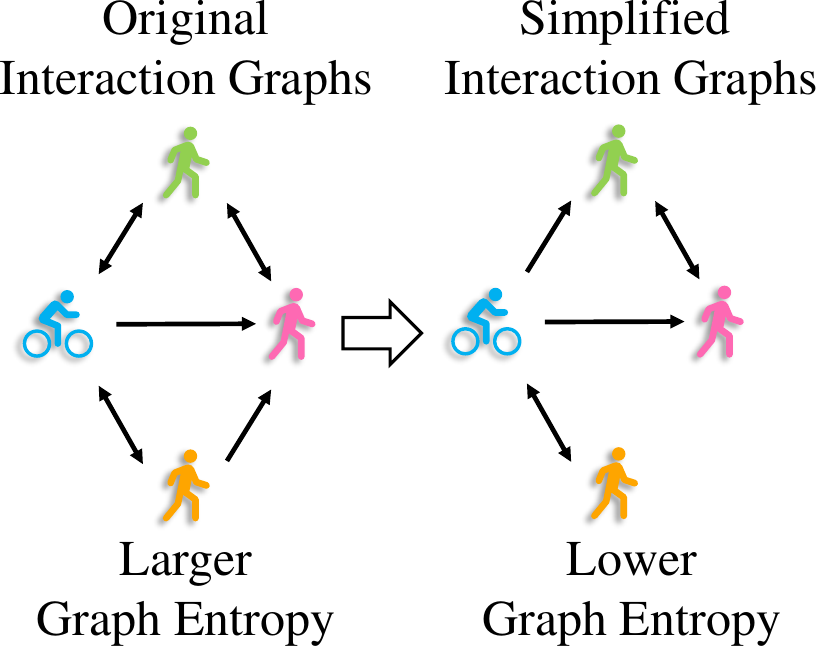}}
  \label{fig:graph_entropy}
  \hfil
  \subfloat[Mixup training strategy.]{\includegraphics[width=0.4\textwidth]{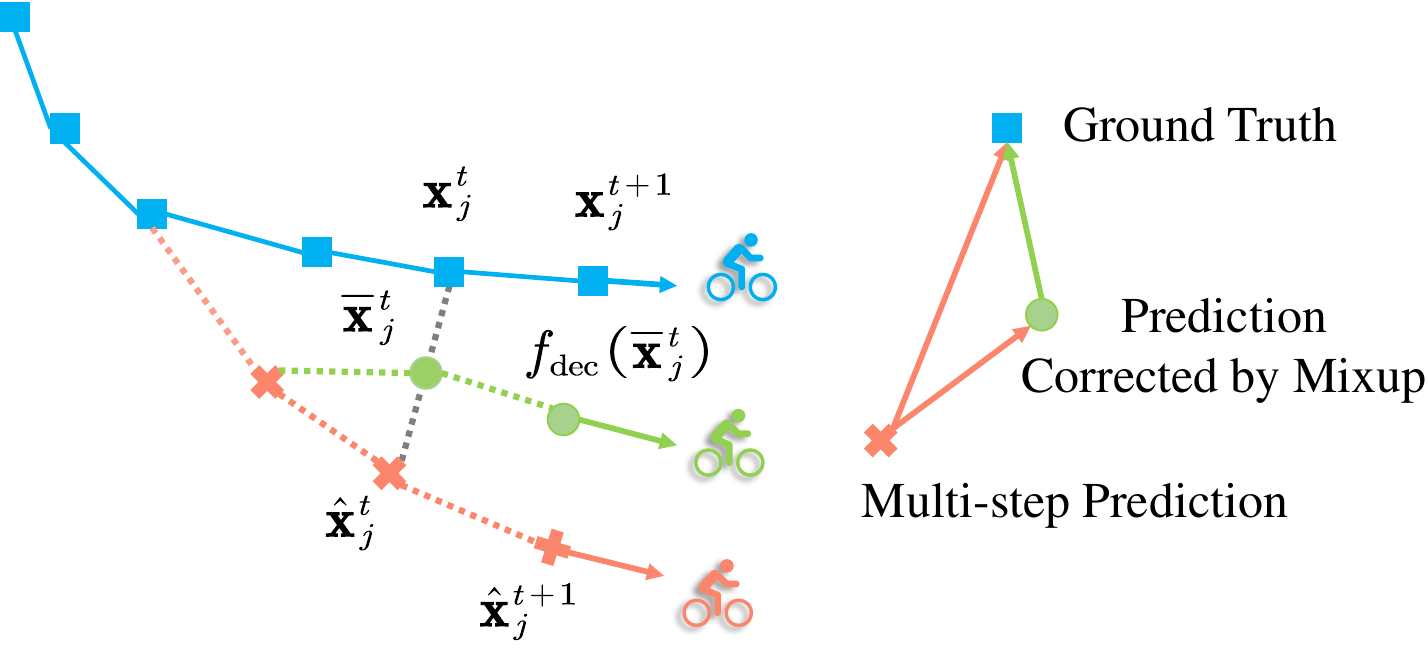}}
  \label{fig:mixup}
  \caption{(a) Visualization of the error sources from both the spatial and temporal perspectives. (b) A lower graph entropy favors a simplified interaction graph. (c) The predicted position and the ground truth position at time $t$ are mixed up to correct the multi-step prediction.}
  \label{fig:error_analysis}
\end{figure*}
The decoder defined in Section~\ref{sec:decoder} makes multi-step predictions recursively, which can be viewed as the forward pass of a multi-layer graph neural network. Fey \textit{et al.}~\cite{fey2021gnnautoscale} showed that under some mild assumption, the errors of the node embeddings grow exponentially on the Lipschitz constants of the GNN model as well as the node degrees with respect to the number of layers. As shown in Fig.~\ref{fig:error_analysis}(a), this result can be intuitively understood in the trajectory prediction problem, since the errors from interacting agents will propagate spatially and accumulate over time. Particularlly, the spatially propagated error is a unique challenge in spatio-temporal time series forecasting like trajectory prediction, which is not covered by traditional research focusing on temporally accumulated error.

The number of agents that affect a target agent can be termed as the in-degree of a node. The distribution of the in-degrees can be imbalanced within an interaction graph, and varies over different graphs. Directly restricting the in-degrees may lead to a degenerate interaction graph that hurts the expressiveness of the model. Instead, this paper turns to the graph entropy~\cite{dehmer2011history}, a global measure for the complexity of graphs with wide applications in sociology, chemistry, etc. It also receives increasing interest in graph machine learning for deciding the node embedding dimension~\cite{luo2021graph} and designing graph pooling modules~\cite{wu2022structural}.

The graph entropy is originally defined on undirected graphs~\cite{dehmer2011history} by introducing a node distribution. The definition can be naturally extended to our case for directed graphs by considering the in-degrees~\cite{dehmer2011history}. Let $d^m_j=\sum^{N}_{i=1}z^m_{ij}$ denote the in-degree of node $j$, and $|\mathcal{E}^m|=\sum_{i\neq j}z^m_{ij}$ denote the total number of edges. Then, a graph entropy is defined as follows,
\begin{equation}
  \text{H}_{\mathcal{G}}[\mathbf{Z}^m]= -\frac{1}{\ln N}{\textstyle\sum^{N}_{j=1}(d^m_{j}/|\mathcal{E}^m|)\ln (d^m_{j}/|\mathcal{E}^m|),}
  \label{eq:graph_entropy}
\end{equation}
where $\ln N$ is a normalization constant for the entropy. The graph entropy weighted by a nonnegative coefficient $\gamma$ is added to the original loss function to penalize the graph complexity,
\begin{equation}
  \mathcal{L}'=\mathcal{L}+\frac{\gamma}{M} {\textstyle\sum^M_{m=1}\text{H}_{\mathcal{G}}[\mathbf{Z}^m]}.
  \label{eq:error_plus_entropy}
\end{equation}
Since $\mathcal{L}$ represents the average prediction error of a single agent per step and $\text{H}_{\mathcal{G}}[\mathbf{Z}^m]$ lies in $[0, 1]$, $\gamma$ stands for the strength of graph complexity penalization to a single node. Eq.~(\ref{eq:error_plus_entropy}) constrains the latent graphs from the probabilistic perspective, and is a meaningful extension of the entropy regularization to graphs, an important class of irregular data.

Ignoring the reconstruction error, the graph entropy is maximized when all in-degrees are equal. The characterization of the minimizer is trickier. For simplicity, it is informally stated as the following theorem.
\begin{theorem}[\text{\textbf{Informal}}]
  Given the numbers of nodes and edges, the graph entropy is minimized when the edges are centered on a few nodes.
  \label{prop:minimizer_graph_entropy}
\end{theorem}
Obviously, the minimizer given by Theorem~\ref{prop:minimizer_graph_entropy} can be rarely reached when considering the reconstruction error. This paper derives a more practical corollary that helps us understand the optima.
\begin{corollary}
Given the number of nodes, a graph with
fewer edges has a smaller lower bound of graph entropy.
\label{coro:entropy_equal_density}
\end{corollary}

The corollary points out that the graph entropy and the sparsity are different metrics for graph complexity. A graph with lower graph entropy is not necessarily sparser, while a sparser graph has a potentially lower graph entropy. An empirical study of graph entropy with sparsity constraints is presented in Section~\ref{sec:app_other_graph_constraint}. A formal statement of Theorem~\ref{prop:minimizer_graph_entropy} and the proofs for Theorem~\ref{prop:minimizer_graph_entropy} and Corollary~\ref{coro:entropy_equal_density} using the majorization theory~\cite{marshall1979inequalities} are shown in \textit{Appendix~B}.

\subsection{Mixup Training Strategy}\label{sec:mixup}
To reduce the temporally accumulated errors, this subsection introduces the mixup training strategy. Mixup is a simple and data-agnostic data augmentation trick~\cite{zhang2018mixup}. It is originally proposed to improve the generalization of deep neural networks by training the models on the linear interpolation (convex combination) of two random training samples and their labels. Liang~\textit{et al.}\cite{liang2020simaug} employed mixup to generalize from simulation datasets to real-world datasets in trajectory prediction. Apart from generalization, mixup can also be used to correct the predicted values. As shown in Fig.~\ref{fig:error_analysis}(c), mixing up the true position and the predicted position yields a more accurate prediction. This motivates us to reduce the accumulated error via mixup. 

Given the true position $\mathbf{X}^t$ and its predicted value $\widehat{\mathbf{X}}^t$, a convex combination of them is defined as
\begin{equation}
  \bar{\mathbf{X}}^t=\text{Mix}_\lambda\left(\widehat{\mathbf{X}}^t, \mathbf{X}^t\right)\triangleq\lambda \widehat{\mathbf{X}}^t + (1-\lambda)\mathbf{X}^t,
\end{equation}
where the mixing coefficient $\lambda\in [0,1]$ is sampled from a beta distribution $\text{Beta}(\alpha,\alpha)$ parameterized by a positive number $\alpha$. 

$\bar{\mathbf{X}}^t$ can be regarded as a correction of the predicted position $\widehat{\mathbf{X}}^t$. Thus, $\bar{\mathbf{X}}^t$ can serve as a more accurate starting point for multi-step prediction, which in turn helps to infer a more accurate interaction graph. Furthermore, by simplifying the notation of multi-step predictions as $\widehat{\mathbf{X}}^{t+1}=f_{\text{dec}}(\widehat{\mathbf{X}}^t)$, this paper conjectures that the gap between $\widehat{\mathbf{X}}^{t+1}$ and $f_{\text{dec}}(\bar{\mathbf{X}}^t)$ is smaller than that between $\widehat{\mathbf{X}}^{t+1}$ and $\mathbf{X}^{t+1}$, which is visualized in Fig.~\ref{fig:error_analysis}(c). Under this assumption, alternatively minimizing two intermediate objectives $\|f_{\text{dec}}(\bar{\mathbf{X}}^t)-\mathbf{X}^{t+1}\|^2_F$ and $\|f_{\text{dec}}(\bar{\mathbf{X}}^t)-\widehat{\mathbf{X}}^{t+1}\|^2_F$ can be easier than directly optimizing the original objective $\|\mathbf{X}^{t+1}-\widehat{\mathbf{X}}^{t+1}\|^2_F$.

However, correcting the prediction at each time step may prevent the model from learning to make multi-step predictions. Without loss of generality, this paper proposes to correct the final prediction of each time window. It can balance the model$'$s ability in single-step and multi-step prediction instead of focusing on single-step prediction like teacher forcing. Let $f_{\text{dec}}^n(\bar{\mathbf{X}}^t)$ denotes the $n$-th step prediction from time $t$. Then, the mixup training strategy can be briefly described in the decoding procedure within a time window. The corresponding pseudo code is shown in Algorithm~\ref{alg:mixup}.

\begin{algorithm}[t]
  \caption{Mixup Training Strategy}\label{alg:mixup}
  \LinesNumbered
  \label{pseudo}
  \KwIn{decoder $f_{\text{dec}}$ with the parameter $\theta$, hyperparameter $\alpha$ for the beta distribution, optimization algorithm $A$.
  }
  
  \KwOut{decoder $f_{\text{dec}}$.}
  
  Initialize $\theta$ with random weights;

  \tcc{First update.}
      Compute $\widehat{\mathbf{X}}^{t}$;

      Sample $\lambda\sim\text{Beta}(\alpha, \alpha)$;

      Compute $\bar{\mathbf{X}}^t=\text{Mix}_\lambda(\text{StopGrad}(\widehat{\mathbf{X}}^t),\mathbf{X}^t)$;

      Compute $\mathcal{L}_1=\sum^\tau_{n=1}\left\|\mathbf{X}^{t+n} - f_{\text{dec}}^n(\bar{\mathbf{X}}^t)\right\|^2_F$;

      $\theta := A\left(\mathcal{L}_1;\theta\right)$;

      \tcc{Second update.}

      Compute $\mathcal{L}_2=\sum^\tau_{n=1}\left\|f_{\text{dec}}^n(\widehat{\mathbf{X}}^t) - \text{StopGrad}(f_{\text{dec}}^n(\bar{\mathbf{X}}^t))\right\|^2_F$;

      $\theta := A\left(\mathcal{L}_2;\theta\right)$;

  \Return{decoder $f_{\text{dec}}$.}
  
  \end{algorithm}
$\text{StopGrad}(\cdot)$ in line 4 of Algorithm~\ref{alg:mixup} is a function that treats $\widehat{\mathbf{X}}^t$ as a constant without back-propagation. The parameter $\alpha$ decays over the training epochs, resulting in a gradually flattened beta distribution~\cite{zhang2018mixup} that allows more combinations of the predicted positions and the ground truth. This increases the difficulties in training and encourages the model to make more accurate multi-step predictions. The graph entropy introduced in Section~\ref{sec:graph_entropy} can be added to the loss function to further reduce the accumulated error.

Note that the mixup training strategy is fundamentally different from Zhang~\textit{et al.}$'$s method~\cite{zhang2019bridging}. Our method is designed for the continuous domain while theirs is limited to the discrete domain. As for the key techniques, the ground truth is combined with the predicted value via interpolation rather than random substitution. Moreover, an alternating optimization procedure is tailored for our problem instead of minimizing the original objective.

Finally, this paper theoretically analyzes the effectiveness of the proposed optimization algorithm under some simplified assumptions~\cite{venkatraman2017training}, summarized in the following theorem.
\begin{theorem}
  Let $f$ be a learned model with bounded single-step prediction error $\|f(\mathbf{X}^t)-\mathbf{X}^{t+1}\|\leq \epsilon$. Assume that $f$ is Lipschitz continuous with constant $L_f$ under the norm $\|\cdot\|$. Then, the following upper bounds hold,
  \begin{itemize}
    \item[(1)] $\|f^n(\widehat{\mathbf{X}}^t)-\mathbf{X}^{t+n}\|\leq L_f^n\|\widehat{\mathbf{X}}^t-\mathbf{X}^t\|+\frac{L_f^n-1}{L_f-1}\epsilon$;
    \item[(2)] $\underset{\lambda}{\mathbb{E}}[\|f^n(\bar{\mathbf{X}}^t)-\mathbf{X}^{t+n}\|]\leq \frac{1}{2}L_f^n\|\widehat{\mathbf{X}}^t-\mathbf{X}^t\|+\frac{L_f^n-1}{L_f-1}\epsilon$;
    \item[(3)] $\underset{\lambda}{\mathbb{E}}[\|f^n(\widehat{\mathbf{X}}^t)-f^n(\bar{\mathbf{X}}^t)\|]\leq \frac{1}{2}L_f^n\|\widehat{\mathbf{X}}^t-\mathbf{X}^t\|$.
  \end{itemize}
   
\end{theorem}
When $L_f>1$, the accumulated error grows exponentially w.r.t. the steps, which agrees with empirical results. The two intermediate losses in the mixup training strategy enjoy smaller upper bounds than the original loss, which helps alleviate the error accumulation problem in the training stage. The proof of this theorem is shown in \textit{Appendix~C}.

\subsection{Characteristics of HIMRAE}
\label{sec:charateristics}
HIMRAE is characterized by the following features.
\begin{itemize}
  \item It learns edge-featured interaction graphs, where a directed edge indicates the existence of interaction and an edge feature implicitly decides the interacting effect. It does not rely on pre-defined adjacency matrices or edge types that may fail to describe heterogeneous interactions precisely.
  \item The heterogeneous attention mechanism can discriminate the importance of different types of agents, requiring only linear space complexity.
  \item The graph entropy, an analogy of Shannon entropy on graphs, reduces the spatially propagated error by penalizing the graph complexity. Unlike sparsity constraints, it controls the graph complexity appropriately without learning an over sparse graph that hurts the prediction precision.
  \item The mixup training strategy can balance the model$'$s ability of single-step and multi-step prediction, and thereby reduce the temporally accumulated error. Unlike teacher forcing, it can effectively narrow the gap between training and testing.
  \item Both graph entropy and mixup training strategy have certain theoretical justifications that can guide the hyperparameter selection and help with experimental analysis.
\end{itemize}

\section{Experiments}
\subsection{Datasets}
\begin{table}[b]
  \renewcommand{\arraystretch}{1.3}
  \caption{Statistics of Datasets.}
  \centering
    \begin{tabular}{c|c|c|c}
    \hline
    Datasets & NBA   & H3D   & SDD \\
    \hline
    \# Scenes & 100   & 121   & 56 \\
    \# Categories & 3     & 8     & 6 \\
    \# Agents & 11    & [4, 30] & [5, 61] \\
    \# Samples & 10k      & 10k      & 27k \\
    Sampling rate & 2.5Hz & 2.5Hz & 2.5Hz \\
    Length unit & Meter & Meter & Pixel \\ 
    Numerical range of $x$ & [0, 28.65] & [-16, 40] & [0, 1960] \\
    Numerical range of $y$ & [0, 15.24] & [-40, 40] & [0, 1977] \\
    Historical steps/Future steps & 5/10  & 5/10  & 8/12 \\
    \hline
    \end{tabular}%
  \label{tab:statistics}%
\end{table}%

Following Li \textit{et al.}~\cite{li2020evolvegraph}, the proposed method is evaluated on three real-world datasets involving heterogeneous agents, NBA SportVU Dataset (NBA), Honda 3D Dataset~\cite{patil2019h3d} and Stanford Drone Dataset (SDD)~\cite{robicquet2016learning}, described as follows.
\begin{itemize}
  \item NBA: a trajectory dataset collected by NBA with the SportVU tracking system, containing the trajectories of the ball and ten players, with five from the home team and the rest from the visiting team.
  \item H3D: a large-scale full-surround 3D multi-object detection and tracking dataset, which provides the trajectories of eight types of traffic participants, including pedestrians, animals, cyclists, motorcyclists, cars, buses, trucks, and other vehicles.
  \item SDD: a university campus trajectory dataset, containing the trajectories of six types of traffic participants, pedestrians, cars, buses, bikers, skaters and carts.
\end{itemize}

The raw data are preprocessed by ourselves according to the original paper~\cite{li2020evolvegraph} with our best efforts. Details of the data preprocessing are described in \textit{Appendix~D}. Some important statistics of the datasets are listed in Table~\ref{tab:statistics}.
\subsection{Evaluation Metrics}
Following previous works~\cite{alahi2016social,huang2019stgat}, the average displacement error (ADE) and the final displacement error (FDE) are used as the evaluation metrics. The ADE and FDE for the predicted trajectory of a single agent are defined as follows. 
\begin{itemize}
  \item Average displacement error: the average Euclidean distance between the predicted positions and the ground truth positions over all future steps.
  \item Final displacement error: the Euclidean distance between the predicted position and the ground truth position at the final time step.
\end{itemize}

To compare the performance of different generative models, $20$ samples are drawn, and both the minimum and the mean of ADE and FDE are calculated. Most of the existing works~\cite{robicquet2016learning, huang2019stgat, mohamed2020social, li2020evolvegraph} are evaluated in minimum ADE and FDE, which may be unreliable in real-world applications since little prior knowledge is available to select the best prediction~\cite{ivanovic2019trajectron}. Therefore, the mean ADE and FDE can serve as important complemental metrics that measure the average performance of generative models.

\subsection{Baselines}
Our method is compared with four GNN based interaction modeling methods to validate its effectiveness.
\begin{itemize}
  \item STGAT \cite{huang2019stgat}: this method applies the graph attention network to a fully connected graph to model the interactions among pedestrians.
  \item Social-STGCNN~\cite{mohamed2020social}: this method defines a distance-based weighted interaction graph at each time step, and applies a spatio-temporal graph convolutional neural networks to capture the spatio-temporal dynamics of pedestrians.
  \item STAR~\cite{yu2020spatio}: this method defines interaction graphs for pedestrians by distance thresholding, and adopts the Transformer for both spatial and temporal modeling.
  \item EvolveGraph~\cite{li2020evolvegraph}: this method infers dynamic multi-relational interaction graphs to model interactions among heterogeneous agents.
\end{itemize}

The codes of STGAT\footnote{https://github.com/huang-xx/STGAT}, Social-STGCNN\footnote{https://github.com/abduallahmohamed/Social-STGCNN} and STAR\footnote{https://github.com/Majiker/STAR} are public and thus directly used in our experiments. EvolveGraph is coded by ourselves according to the original paper. Implementation details of our method can be found in \textit{Appendix~D}. 

Some variants of our method are introduced as follows to demonstrate the effectiveness of the heterogeneous attention mechanism, the graph entropy and the mixup training strategy.
\begin{itemize}
  \item HIMRAE${}_{\text{HOMO}}$: this variant replaces the category-aware mappings of HAM with an identity mapping.
  \item HIMRAE w/ GE: a variant trained only with the graph entropy.
  \item HIMRAE w/ mixup: a variant trained only with the mixup training strategy.
  \item HIMRAE w/ GE \& mixup: a variant trained with both the graph entropy and the mixup training strategy.
\end{itemize}
\subsection{Comparison with Baselines}
\begin{table*}[t]
  \renewcommand{\arraystretch}{1.3}
  \centering
  \caption{ADEs/FDEs of trajectory prediction on different datasets. The minimum and the mean of the metrics calculated on $20$ random samples are reported. For all evaluation metrics, lower is better. The values may differ from those reported by Li~\textit{et al.}~\cite{li2020evolvegraph} since the experiments are rerun on all datasets using our data preprocessing. Results highlighted with * are statistically significant with $p<0.01$ by comparing with EvolveGraph, the best baseline overall. Details of the $p$-values can be found in Appendix~D.}
    \begin{tabular}{c|c|c|c|c|c|c}
    \hline
    \multicolumn{1}{c|}{Datasets} & \multicolumn{2}{c|}{NBA} & \multicolumn{2}{c|}{H3D} & \multicolumn{2}{c}{SDD} \\
    \hline
    {Units} & \multicolumn{2}{c|}{Meters} & \multicolumn{2}{c|}{Meters} & \multicolumn{2}{c}{Pixels}\\
    \hline
    Metrics & \multicolumn{1}{c|}{min${}_{20}\downarrow$} & \multicolumn{1}{c|}{mean${}_{20}\downarrow$} & \multicolumn{1}{c|}{min${}_{20}\downarrow$} & \multicolumn{1}{c|}{mean${}_{20}\downarrow$} & \multicolumn{1}{c|}{min${}_{20}\downarrow$} & \multicolumn{1}{c}{mean${}_{20}\downarrow$} \\
    \hline
    STGAT & \textbf{0.07}/\textbf{0.13}      & 0.20/0.42      & 0.74/1.45      & 3.34/8.34      & \textbf{12.7}/\textbf{23.3}       & 27.9/58.8 \\
    Social-STGCNN &0.10/0.18 	&0.18/0.34 	&1.51/2.46 	&2.61/5.01 	&21.2/32.7 	&40.5/73.6   \\
    STAR  &0.08/0.18 	&0.25/0.62 	&0.90/2.09 	&2.56/6.92 	&15.9/32.4 	&31.1/70.6  \\
    EvolveGraph &0.12/0.21 	&0.19/0.34  &0.58/1.00 	&0.81/1.49 	&20.9/39.8 	&28.3/54.0  \\
    \hline
    HIMRAE w/ TF &0.55/0.72 	&1.75/2.92 		&5.43/10.17 	&12.05/25.81 		&178.0/332.4 	&315.8/592.9   \\
    HIMRAE w/ TF+ &0.11/0.20 	&0.19/0.36 		&0.63/1.12 	&1.04/2.08 		&20.7/36.9 	&27.5/51.7   \\
    \hline
    HIMRAE${}_{\text{HOMO}}$ &0.09/0.16 	&0.17/0.31 &0.41/0.60 	&0.70/1.32 &17.2/31.2 	&29.2/54.9  \\
    HIMRAE &0.09/0.16 	&0.16/0.29 &0.40/0.57 &0.69/1.28 &16.9/30.5 	&29.0/54.3	  \\
    HIMRAE w/ GE  &0.09/0.16 	&0.16/0.29 &0.37/0.54 	&0.64/1.22 &18.0/33.6 	&26.8/51.2  \\
    HIMRAE w/ mixup &0.09/0.18 	&\textbf{0.14}/\textbf{0.28} &0.36/0.54 	&0.58/1.04  &19.3/37.5 	&\textbf{25.2}/49.9 \\
    \hline
    HIMRAE w/ GE \& mixup  &0.09${}^*$/0.17${}^*$ 	&0.15${}^*$/\textbf{0.28}${}^*$ &\textbf{0.35}${}^*$/\textbf{0.50}${}^*$ 	&\textbf{0.56}${}^*$/\textbf{0.99}${}^*$ &18.8${}^*$/36.4${}^*$ 	&\textbf{25.2}${}^*$/\textbf{49.4}${}^*$  \\
    \hline
    \end{tabular}%
  \label{tab:baselines}%
\end{table*}%
As shown in Table~\ref{tab:baselines}, our method HIMRAE w/ GE \& mixup achieves the lowest mean ADEs/FDEs on all datasets, and the lowest minimum ADEs/FDEs on the H3D dataset. The overall results validate the superiority of the proposed method.

All methods perform well on the NBA dataset, where most agents (players) are of the same type except the ball. STGAT achieves the lowest minimum ADEs/FDEs, while its mean ADEs and FDEs are larger than other baselines. STGAT is trained on the variety loss~\cite{huang2019stgat}, i.e., evaluating the training batch several times and back-propagating through the lowest loss. This training strategy encourages STGAT to make diversified predictions with a possible large variance, leading to relatively poor performance on average. The problem becomes more severe on heterogeneous datasets like H3D and SDD. Our method outperforms other baselines on the NBA dataset since the category information is considered and the two training strategies further reduce the errors. Lower mean ADEs/FDEs show that our method can make more accurate predictions with lower variance, which is more reliable in practical scenarios.

On the H3D dataset, both our method and EvolveGraph achieve significantly lower predicting errors than the rest methods that are originally designed for pedestrian trajectory prediction. The results show that heterogeneous agents present different behavior patterns, which can not be effectively captured by homogeneous modeling. EvolveGraph, a baseline that considers both heterogeneous interaction modeling and decoding, make even less accurate predictions than HIMRAE. Maybe the heterogeneous attention mechanism in our method can distinguish the importance of heterogeneous neighbors, while EvolveGraph assigns equal weights to all neighbors in the decoder.

On the SDD dataset, methods designed for homogeneous systems achieve comparable predicting errors as our method and EvolveGraph. A possible reason is that the dominant class of agents on the SDD dataset is the pedestrian. Compared with EvolveGraph, our method achieves significantly lower mean ADEs/FDEs, because the graph entropy and the mixup training strategy can effectively reduce the accumulated error.

The predicted trajectories of HIMRAE w/ GE \& mixup and the most related baseline EvolveGraph are visualized in \textit{Appendix~D} for qualitative analysis.
\subsection{Ablation Study}
\subsubsection{Effect of Heterogeneous Attention Mechanism}

\begin{figure}[!t]
  \centering
  \subfloat[Distance based adjacency matrices at $t=8$ and $t=13$.]{\includegraphics[width=0.3\columnwidth]{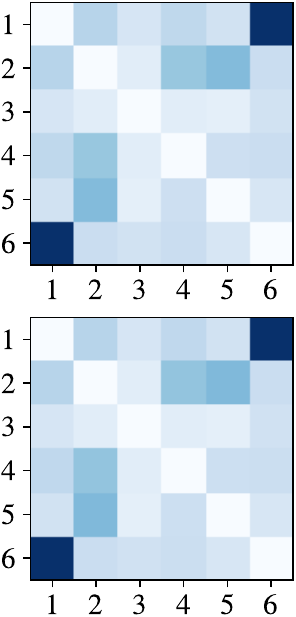}}
  \label{fig:distance}
  \hfil
  \subfloat[{Interacting probabilities for $t\in [6, 10]$ and $t\in [11,15]$.}]{\includegraphics[width=0.3\columnwidth]{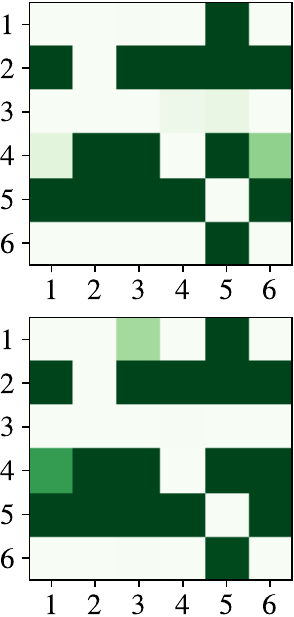}}
  \label{fig:relation}
  \hfil
  \subfloat[Attention weights at $t=8$ and $t=13$.]{\includegraphics[width=0.3\columnwidth]{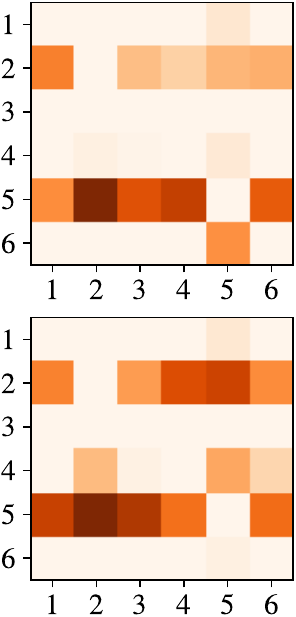}}
  \label{fig:attention}
  \hfil
  \subfloat[Trajectories of a six-agent system.]{\includegraphics[width=0.75\columnwidth]{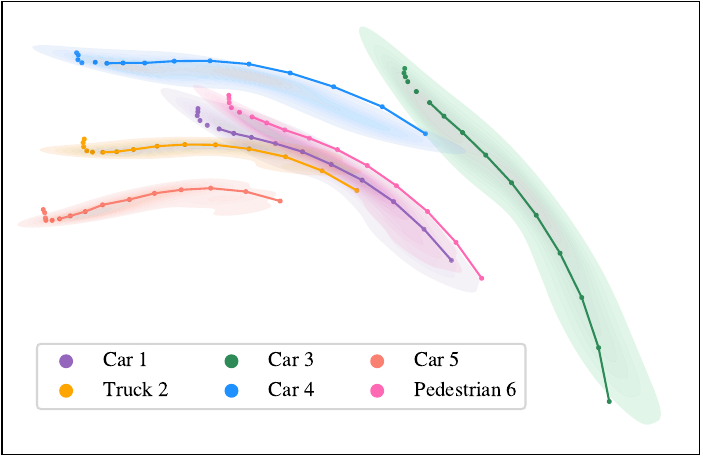}}
  \label{fig:trajectory}
  \label{fig:ham}
  \caption{For matrices in Subfig.~(a)-(c), lighter indicates stronger connections.  In Subfig.~(d), historical trajectories are in dots, ground truth trajectories to be predicted are in solid lines, while the predicted trajectories are visualized using the kernel density estimation.}
  \label{fig:att_traj}
\end{figure}
As shown in Table~\ref{tab:baselines}, HIMRAE consistently outperforms HIMRAE${}_{\text{HOMO}}$ on H3D and SDD datasets, while their performances are comparable on the NBA dataset. The results validate that the category-aware mappings are key components of HAM, which extends the normal attention mechanism to handle the interactions among heterogeneous agents. 

To gain an intuitive understanding of HAM, this paper visualizes the distance-based adjacency matrices, the probability matrices of the interacting relations, the attention weights, and the predicted trajectories in Fig~\ref{fig:att_traj}. Compared with the distance-based adjacency matrices, the inferred interaction graph can capture asymmetric relations between agents. For example, during $6\leq t \leq 10$, Car 1 moves away from Truck 2, while Truck 2 is not affected by Car 1. During the same period, although the interacting graph tells Car 5 is affected by almost all other agents except Car 3, HAM helps to identify Pedestrian 6 as the most important agent to avoid. Similar results can be found for $t\in [11,15]$. From this case, one finds that the interaction graph within a time window describes the coarsened interactions, while the attention weights describe the fine-grained interactions at each time step.

\subsubsection{Effect of Graph Entropy}
\begin{figure*}[!t]
  \centering
  \subfloat[Case I: Equal density with smaller entropy.]{\includegraphics[width=0.32\textwidth]{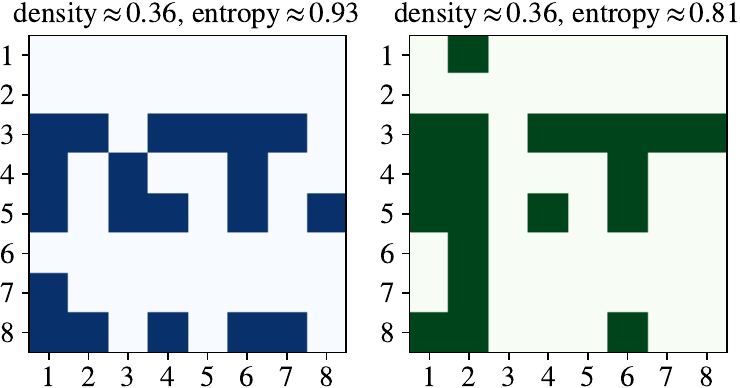}}
  \label{fig:equal_density}
  \hfil
  \subfloat[Case II: Larger density with smaller entropy.]{\includegraphics[width=0.32\textwidth]{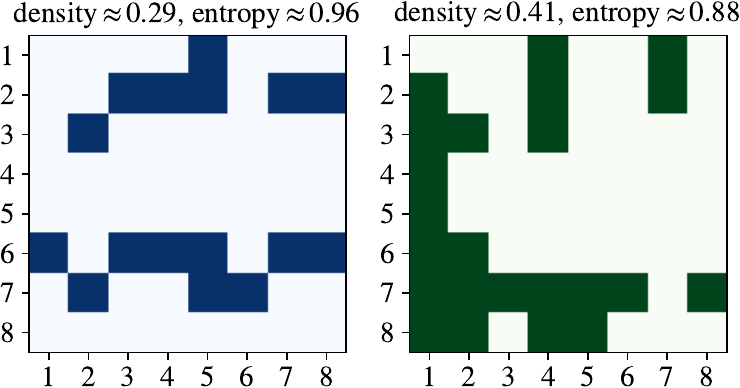}}
  \label{fig:smaller_density}
  \hfil
  \subfloat[Case III: Smaller density with smaller entropy.]{\includegraphics[width=0.32\textwidth]{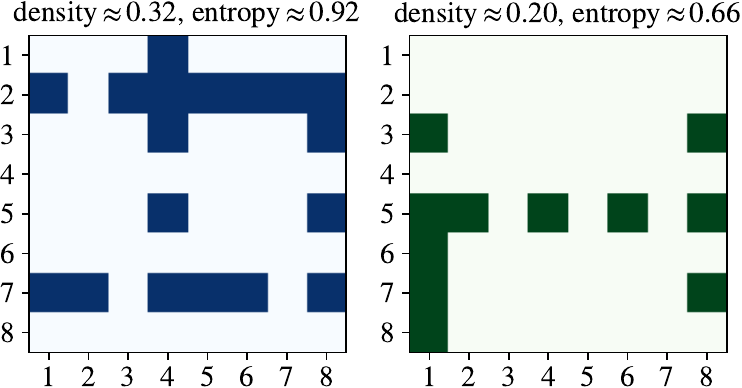}}
  \label{fig:larger_density}
  \caption{{Reduction of graph entropy at different graph density. Graphs inferred by HIMRAE and HIMRAE w/ GE are in blue and green, respectively.}}
  \label{fig:density_vs_entropy}
\end{figure*}
\begin{figure*}
  \begin{minipage}{0.32\textwidth}
    \centering
  \includegraphics[width=\columnwidth]{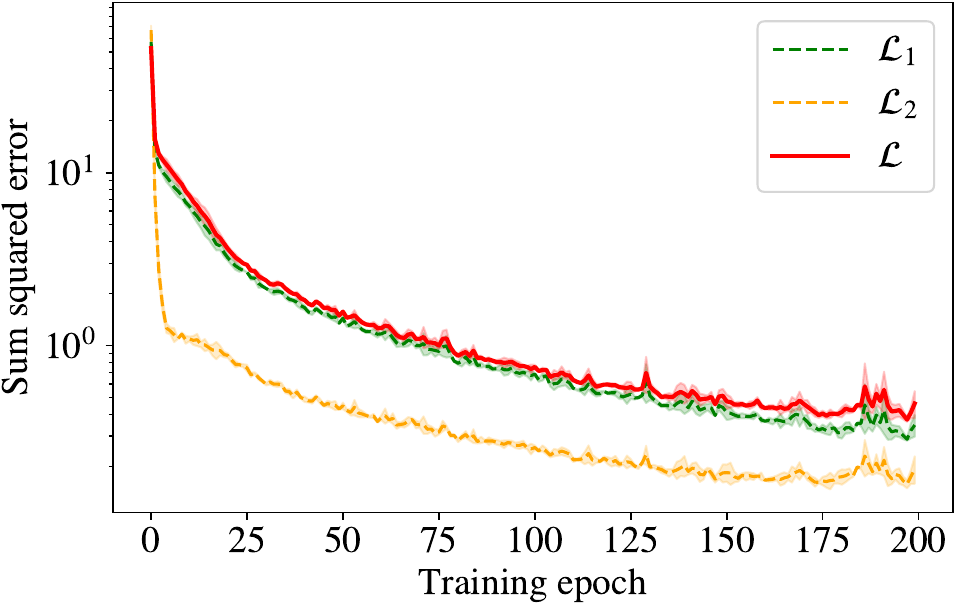}
  \caption{{Training loss of mixup strategy.}}
  \label{fig:mixup_orig_first_second}
  \end{minipage}
  \hfil
  \begin{minipage}{0.32\textwidth}
    \centering
  \includegraphics[width=\columnwidth]{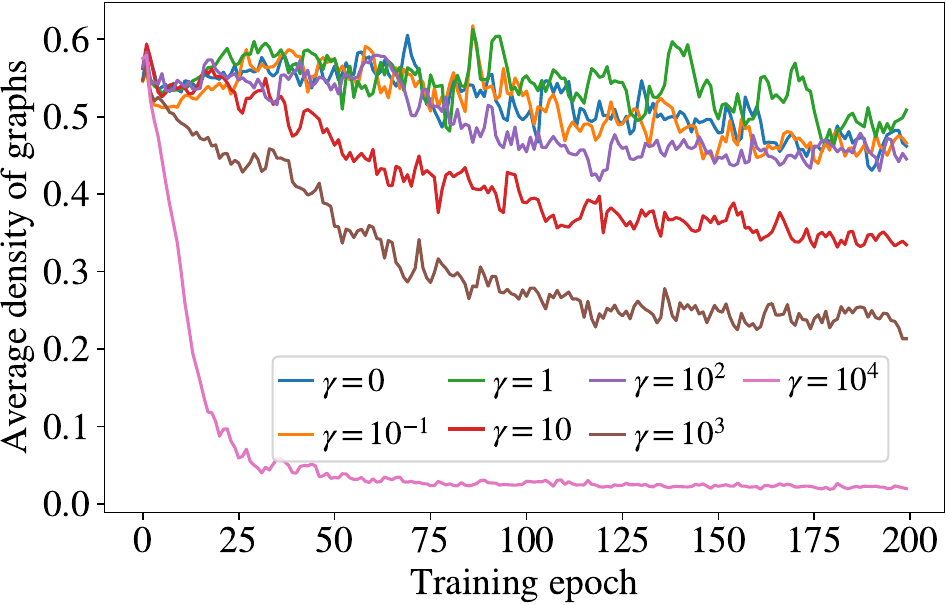}
  \caption{{Density versus $\gamma$.}}
  \label{fig:density_vs_gamma}
  \end{minipage}
  \hfil
  \begin{minipage}{0.32\textwidth}
    \centering
  \includegraphics[width=\columnwidth]{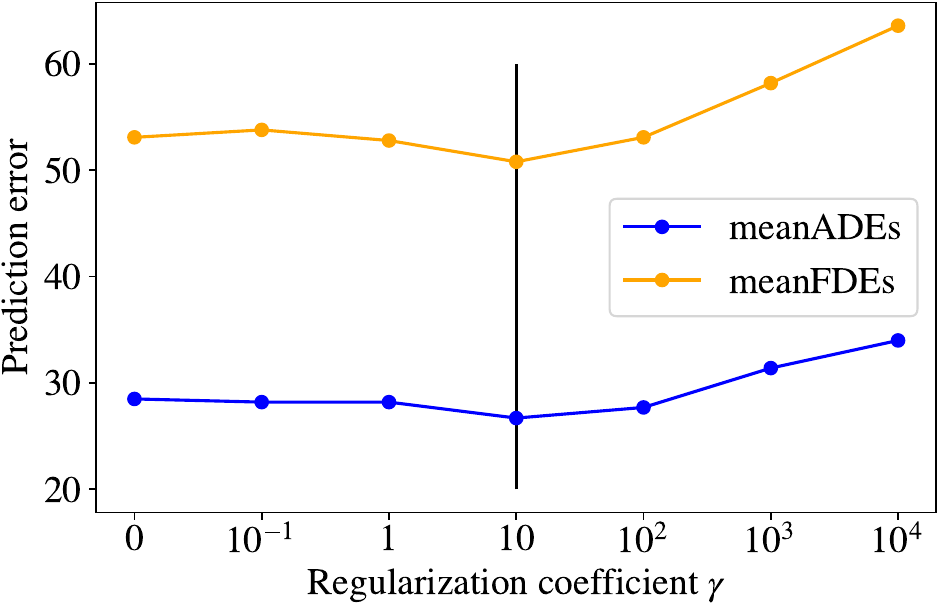}
  \caption{{Prediction error versus $\gamma$.}}
  \label{fig:error_vs_gamma}
  \end{minipage}
\end{figure*}

\begin{table}[!t]
  \renewcommand{\arraystretch}{1.3}
  \caption{Average Graph Entropy and Density on Different Datasets.}
  \centering
  \resizebox{\columnwidth}{!}{
  \begin{tabular}{c|c|c|c|c|c|c}
    \hline
    \multicolumn{1}{c|}{Datasets} & \multicolumn{2}{c|}{NBA} & \multicolumn{2}{c|}{H3D} & \multicolumn{2}{c}{SDD} \\
    \hline
    Metrics & \multicolumn{1}{c|}{GE} & \multicolumn{1}{c|}{Density} & \multicolumn{1}{c|}{GE} & \multicolumn{1}{c|}{Density} & \multicolumn{1}{c|}{GE} & \multicolumn{1}{c}{Density} \\
    \hline
    HIMRAE &\textbf{0.95} & 0.50	&0.87 & 0.39	&0.90 & 0.57\\
    HIMRAE w/ GE &\textbf{0.95} & \textbf{0.43} &\textbf{0.84} & \textbf{0.38}	&\textbf{0.88} & \textbf{0.56}\\
    \hline
    \end{tabular}%
  }
  \label{tab:avg_graph_entropy}%
\end{table}%

As shown in Table~\ref{tab:baselines}, by adding the graph entropy, the mean ADEs/FDEs decreases on H3D and SDD datasets, while there is no significant difference on the NBA dataset. To see how graph entropy reduces the graph complexity, this paper reports the average graph entropy and average graph density $|\mathcal{E}|/(N(N-1))$ of HIMRAE and HIMRAE w/ GE in Table~\ref{tab:avg_graph_entropy}.

From Table~\ref{tab:avg_graph_entropy}, HIMRAE w/ GE achieves lower graph complexity than HIMRAE on all datasets except that there is no significant difference in graph entropy on the NBA dataset. The results also agree with the prediction errors reported in Table~\ref{tab:baselines}, where graph entropy is more effective on the H3D dataset and SDD dataset. However, as discussed in Section~\ref{sec:graph_entropy}, a graph with lower graph entropy is not necessarily sparser. A case study is conducted on the H3D dataset to illustrate that HIMRAE w/ GE can infer graphs with lower graph entropy despite the graph density. As shown Fig.~\ref{fig:density_vs_entropy}, graphs inferred by HIMRAE w/ GE can achieve a lower graph entropy when the graph densities are equal, larger or smaller, because the node distributions are less spread out.

\subsubsection{Effect of mixup strategy}

As shown in Table~\ref{tab:baselines}, the mixup training strategy consistently reduces the mean ADEs/FDEs of HIMRAE on different datasets. The combination of mixup and the graph entropy can further improve the performance. Nonetheless, the minimum ADEs/FDEs increase on the NBA dataset and the SDD dataset. Maybe there exists a trade-off between minimum errors and mean errors, and the mixup training strategy tends to reduce the mean errors at the cost of increasing the minimum errors.

To explore how mixup eases the training of the model, this paper visualizes the training loss and the two intermediate losses in Fig.~\ref{fig:mixup_orig_first_second}. Besides, the intermediate loss $\mathcal{L}_1$ is slightly lower than the original loss $\mathcal{L}$ and $\mathcal{L}_2$ is much smaller. The result agrees with the theoretical analysis, and verifies that mixup does reduce the accumulated errors in the training stage. Note that $\mathcal{L}_2$ is significantly lower than $\mathcal{L}_1$, which means that imitating the behavior of multi-step predictions based on corrected values is easier than directly learning from the ground truth trajectories. Despite the advantages mentioned above, HIMRAE w/ mixup requires more training time, as it includes extra forward and backward passes in one epoch.

\section{Discussions}
\subsection{Selection of Regularization Coefficient $\gamma$}
The hyperparameter $\gamma$ controls the strengths of complexity regularization. The best $\gamma$ varies with datasets since the units of positions and the dynamics are different. Without loss of generality, this paper illustrates the choice of $\gamma$ by training HIMRAE w/ GE on the SDD dataset with $\gamma$ selected from $\{0,10^{-1},1,10,10^2,10^3,10^4\}$. The average graph densities during the training procedure are shown in Fig.~\ref{fig:density_vs_gamma}, and the mean ADEs/FDEs are shown in Fig.~\ref{fig:error_vs_gamma}. From Fig.~\ref{fig:density_vs_gamma}, a larger $\gamma$ approximately leads to a sparser graph, which agrees with the theoretical results in Corollary~\ref{coro:entropy_equal_density}. When $\gamma=10^4$, the average graph density declines dramatically to zero, resulting in an almost empty graph. Therefore, a suitable $\gamma$ should be chosen from the rest candidates. From Fig.~\ref{fig:error_vs_gamma}, the model reaches the minimum predictor error at $\gamma=10$.
\subsection{Comparing Graph Entropy With Sparsity Constraints}\label{sec:app_other_graph_constraint}
\begin{table}[t]
  \renewcommand{\arraystretch}{1.3}
  \caption{Effect of Different Regularization Terms. }
  \centering
    \begin{tabular}{c|c|c|c}
    \hline
    Metrics & min${}_{20} \downarrow$     &  mean${}_{20} \downarrow$   & Density (\%) \\
    \hline
    HIMRAE &0.40/0.57 &0.69/1.28 & 38.6	  \\
HIMRAE w/ GE  &\textbf{0.37}/\textbf{0.54} 	&\textbf{0.64}/\textbf{1.22} & 37.9 \\
    HIMRAE w/ $R_{\text{density}}$ & 0.40/0.58    & 0.69/1.30   & \textbf{36.0} \\
    HIMRAE w/ $R_{\text{degree}}$ & 0.42/0.62   & 0.71/1.33   & 37.2 \\
    \hline
    \end{tabular}%
  \label{tab:different_regularization}%
\end{table}%

Apart from graph entropy, there are other choices of regularization to penalize the graph complexity like the sparsity constraints. Without loss of generality, this paper considers two simple and differentiable candidates, the density constraint and the maximum node degree constraint, i.e.,
\begin{equation*}
  \begin{aligned}
    R_{\text{density}}(\mathbf{Z}^{1:M})&=\frac{\gamma}{M}{\textstyle\sum^M_{m=1}\frac{1}{N(N-1)}\sum_{i\neq j}z^m_{ij},}\\
  R_{\text{degree}}(\mathbf{Z}^{1:M})&=\frac{\gamma}{M}{\textstyle\sum^M_{m=1}\frac{\max_{j}d^m_{j}}{N}.}
  \end{aligned}
\end{equation*}
The resulting variants of our method are denoted as \textbf{HIMRAE w/ $R_{\text{density}}$} and \textbf{HIMRAE w/ $R_{\text{degree}}$}.

A comparative experiment is conducted on the H3D dataset to show the advantages of graph entropy over the sparsity constraints in trajectory prediction. The regularization coefficients for HIMRAE w/ $R_{\text{density}}$ and HIMRAE w/ $R_{\text{degree}}$ are chosen as $10^{-4}$ and $10^{-4}$ respectively via cross-validation. The minimum ADEs/FDEs, mean ADEs/FDEs, and the average graph density are reported in Table~\ref{tab:different_regularization}.

From Table~\ref{tab:different_regularization}, HIMRAE w/ $R_{\text{density}}$ and HIMRAE w/ $R_{\text{degree}}$ achieve lower densities, while HIMRAE w/ GE achieves the lowest prediction errors. The results demonstrate that sparer graphs do not necessarily lead to lower prediction error. Compared with HIMRAE w/ $R_{\text{density}}$, HIMRAE w/ GE can rationally reduce the graph complexity without learning over sparse graphs that hurt the prediction precision. Compared with HIMRAE w/ $R_{\text{degree}}$, our method can impose a global constraint to the interactions instead of only paying attention to the node with the maximum degree.

\subsection{Comparing mixup Training Strategy With Teacher Forcing}\label{sec:app_vs_teacher_forcing}
\begin{figure*}[t]
  \centering
  \includegraphics[width=\textwidth]{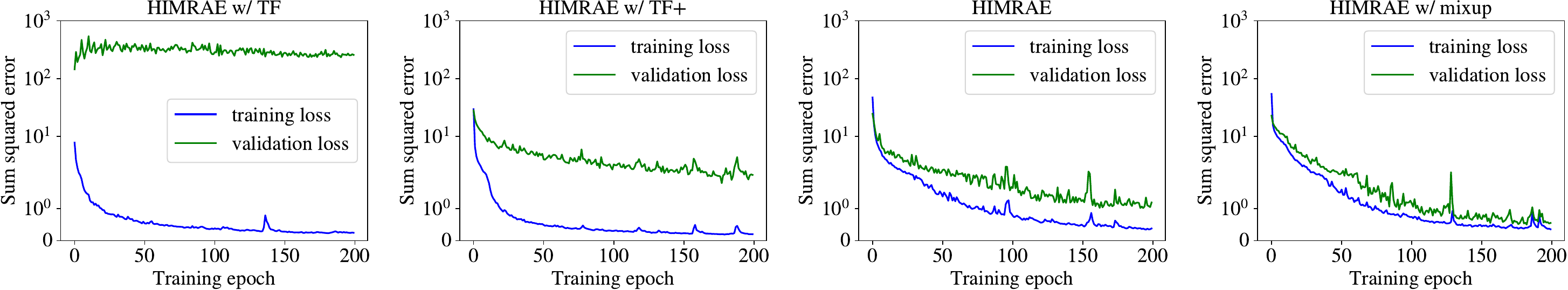}
  \caption{{Training loss and validation loss of different methods.}}
  \label{fig:mixup_vs_teacher_forcing}
\end{figure*}
This subsection compares the performance of mixup training strategy with teacher forcing to show our advantages. Unfortunately, we find teacher forcing fails to train satisfactory models in our dataset (see Table~\ref{tab:baselines} and Fig.~\ref{fig:mixup_vs_teacher_forcing}(a)). For a fair comparison, this paper also considers a variant of teacher forcing that feeds the ground truth positions at every $\tau$ steps. The methods with teacher forcing and its variant are denoted as \textbf{HIMRAE w/ TF} and \textbf{HIMRAE w/ TF+}, respectively. This paper evaluates the prediction errors of teacher forcing and its variant on all datasets, and the results are shown in Table~\ref{tab:baselines}. 

The results show that models trained under teacher forcing fail to make accurate predictions. HIMRAE w/ TF+ performs much better, but on the NBA dataset and H3D dataset, its performance is even worse than HIMRAE, which is trained without the mixup strategy. Therefore, though sharing similar ideas with teacher forcing, the mixup strategy is substantially different. It can better balance the model$'$s ability for both single-step and multi-step prediction.

This paper further visualizes the training loss and validation loss of HIMRAE, HIMRAE w/ mixup, HIMRAE w/ TF and HIMRAE w/ TF+ on the H3D dataset. As shown in Fig.~\ref{fig:mixup_vs_teacher_forcing}, teacher forcing and its variant suffer from a large gap between training loss and validation loss, which is even larger than that of HIMRAE. By contrast, HIMRAE w/ mixup effectively reduces the training-validation gap.

\subsection{Quality of the Interaction Graphs}\label{sec:app_graph_quality}
This subsection makes an attempt to evaluate the quality of the learned interaction graphs. In our model, the edges in a graph are conditionally independent given the historical trajectories. Therefore, it suffices to evaluate the quality of each edge, which indicates the existence of a directed interaction and its effect. Since the interacting effect is encoded by a continuous hidden variable, it is hard to tell its quality. Therefore, this paper tries to check if our model can identify the existence of an edge.

The problem is studied from two aspects:
\begin{itemize}
  \item Necessity of the edges: whether or not the inferred edges contain no redundancy.
  \item Sufficiency of the edges: whether or not the inferred edges contain all useful edges.
\end{itemize}
Two types of tests are introduced to examine the two aspects above, respectively.
\begin{itemize}
  \item Remove an edge at a time to see if the prediction error increases significantly. Edges that do not affect the prediction error are thought redundant.
  \item Add an edge at a time to see if the prediction error decreases significantly. Edges that help reduce the prediction error are thought missing.
\end{itemize}

Let $E, E_1, E_2$ denote the number of inferred edges, the number of redundant edges, and the number of missing edges, respectively. Then, $E - E_1 + E_2$ is an approximate number of all useful edges. Consequently, the redundant rate and the missing rate are defined as follows.
\begin{itemize}
  \item Redundant rate: $\frac{E_1}{E - E_1 + E_2}$.
  \item Missing rate: $\frac{E_2}{E - E_1 + E_2}$.
\end{itemize}
By definition, a lower redundant rate and missing rate indicate higher qualities. This paper evaluates the two metrics on 256 random samples from the testing set of all datasets. For each sample, this paper first infers the interaction graph via a learned model HIMRAE, and then, removes or adds an edge at a time to see if the mean ADE increases significantly. The difference is measured by a statistical significance test over 20 generated trajectories at the level of $p< 0.05$. 

From Table~\ref{tab:quality_of_graphs}, our method can learn satisfactory interaction graphs with the redundant rate and the missing rate lower than 10\%. Still, the missing rate is relatively high on the H3D dataset, indicating that our method needs to consider potentially missing edges for further improvement.

\begin{table}[t]
  \renewcommand{\arraystretch}{1.3}
  \caption{Quality of Inferred Interaction Graphs on Different Datasets.}
  \centering
    \begin{tabular}{r|c|c|c}
    \hline
    Datasets & NBA   & H3D   & SDD \\
    \hline
    Redundant Rate (\%) & 2.34   & 3.11   & 2.91 \\
    Missing Rate (\%) & 2.32   & 7.16   & 1.79 \\
    \hline
    \end{tabular}%
  \label{tab:quality_of_graphs}%
\end{table}%

\subsection{Choices of the Interaction Graphs}\label{sec:app_graph_choices}
\begin{figure*}[!t]
  \centering
  \subfloat[Case I. Maximum graph similarity leads to lower prediction error.]{\includegraphics[width=0.45\textwidth]{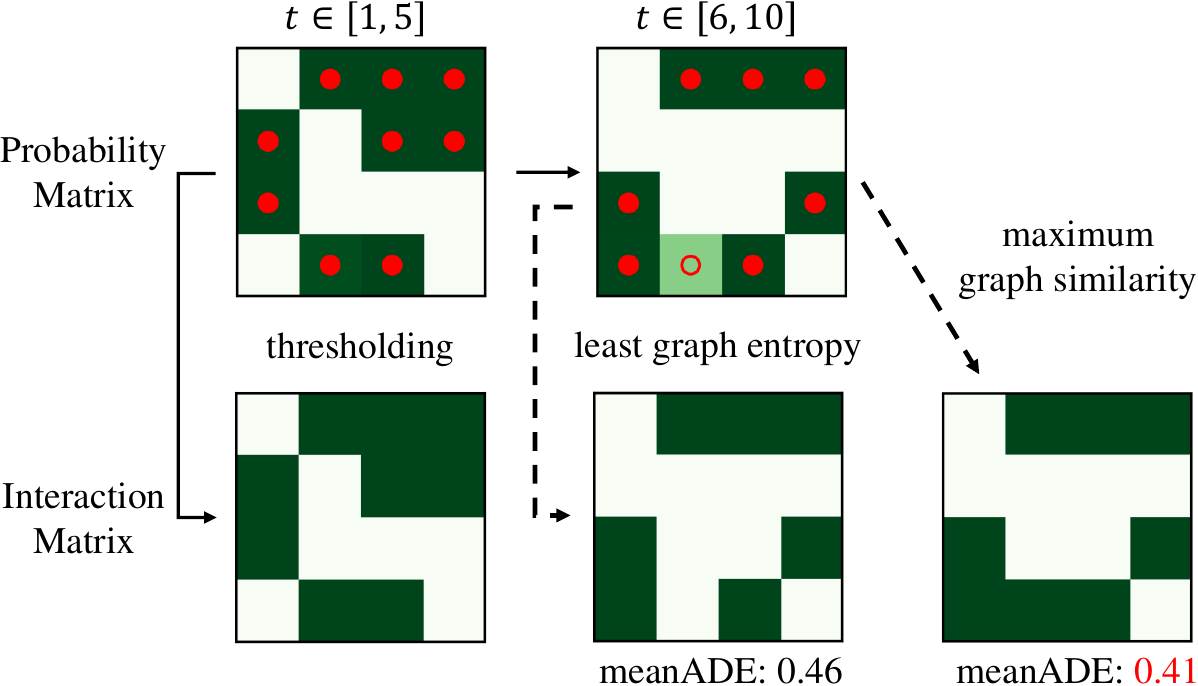}}
  \label{fig:entropy_over_similarity}
  \hfil
  \subfloat[Case II. Least graph entropy leads to lower prediction error.]{\includegraphics[width=0.45\textwidth]{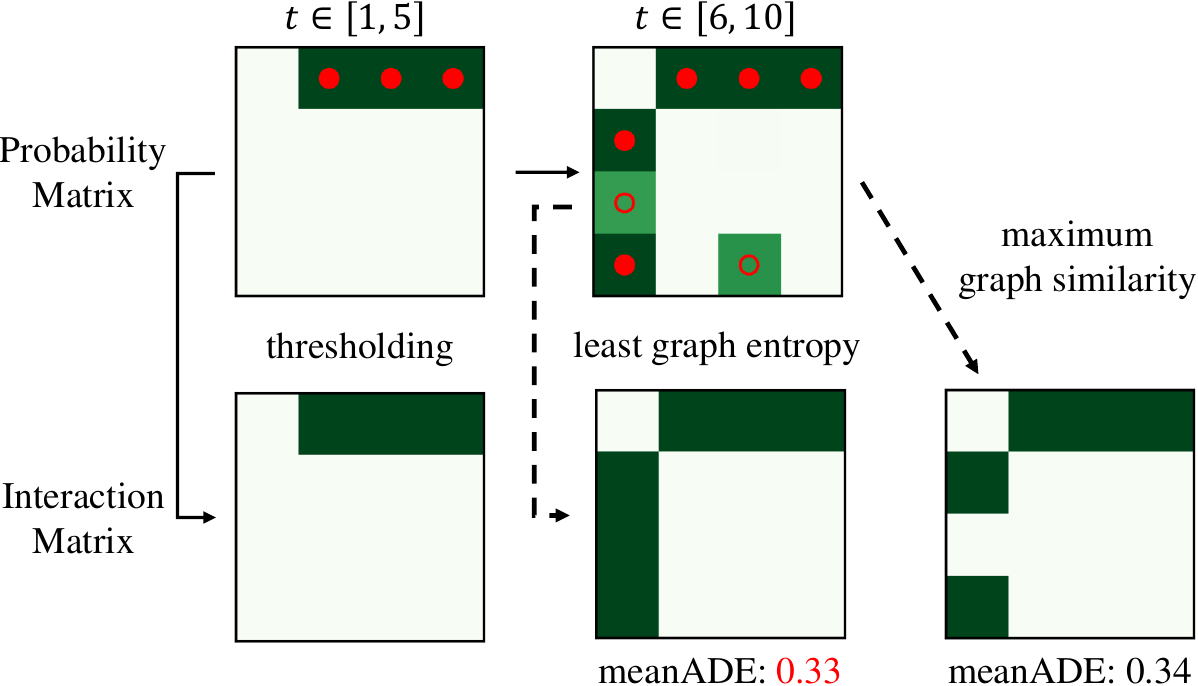}}
  \label{fig:similarity_over_entropy}
  \caption{{Choices of Interaction Graphs. In the probability matrices, red dots stand for probabilities larger than $0.8$, while red circles stand for probabilities in $[0.2, 0.8]$. The lower mean ADEs are highlighted in red.}}
  \label{fig:best_graph}
\end{figure*}
In the experiments, the interaction graphs are sampled randomly to evaluate the performance, while in practical scenarios, users might be interested in how to choose a ``most possible'' graph. First of all, there is generally no gold standard because the ground truth graphs are unobservable. When the future trajectories are available, one can select the graph that minimizes the prediction error. However, when they are unknown, e.g. in the testing stage, an alternative criterion should be chosen before selecting the best graph. 

In our model, edges in a graph are conditionally independent. Once the encoder infers the edge probabilities, the existence of each edge can be decided independently. Edges with small uncertainty are selected deterministically via thresholding, while the rest are selected based on some heuristics. An exemplar procedure is described as follows.
\begin{itemize}
  \item [(1)] Filter out edges with probabilities smaller than a given threshold $\theta_{\text{low}}$.
  \item [(2)] Select edges with probabilities larger than another given threshold $\theta_{\text{high}}$.
  \item [(3)] For edges with probabilities in $[\theta_{\text{low}}, \theta_{\text{high}}]$, this paper adopts the following heuristics.
  \begin{itemize}
    \item[(a)] Select the graph with the least graph entropy.
    \item[(b)] Select the graph that is most consistent with the previous one by measuring the graph similarity, e.g., the $\ell_1$-norm.
  \end{itemize}
\end{itemize}
Other heuristics may be explored based on users$'$ prior knowledge.

Two cases from the H3D dataset are used to illustrate the procedure above. This paper sets $\theta_{\text{low}}=0.2$ and $\theta_{\text{high}}=0.8$ in the experiments. The results are shown in Fig.~\ref{fig:best_graph}. In Case I, heuristics (b) is better than heuristics (a) in terms of prediction error, while the situation reverses in Case II. The case study demonstrates that neither heuristics can ensure the best graph that leads to the lowest reconstruction error.

\section{Conclusion}
This paper proposes an encoder-decoder framework HIMRAE that models the interactions among heterogeneous agents and reduces the accumulated errors of multi-agent trajectory prediction. The encoder can use the historical trajectories to infer dynamic interaction graphs, featured by the interacting relations and corresponding interacting effects. The decoder applies a heterogeneous attention mechanism of linear space complexity to the inferred graphs to model the fine-grained heterogeneous interactions. This paper further figures out the error sources of recursive multi-step prediction from both the spatial and the temporal aspects. The graph entropy is adopted to control the spatially propagated errors and the mixup strategy is proposed to reduce the temporally accumulated errors. Both strategies have theoretical justifications. Extensive experiments on real-world heterogeneous datasets validate the effectiveness of HIMRAE. The results also show the advantages of graph entropy over sparsity constraints, and the advantages of mixup strategy over teacher forcing.

In practical scenarios, the model may be required to predict the trajectories of agents from an unseen category. To tackle this challenge, future work includes using meta-learning to transfer the knowledge of interaction modeling to new agent types. Besides, the interaction graphs inferred by the neural networks lack interpretability. It is worthwhile to explore interpretable interaction graph generation.

\bibliographystyle{IEEEtran}
\bibliography{IEEEabrv,ref}
\includepdf[pages=1-7]{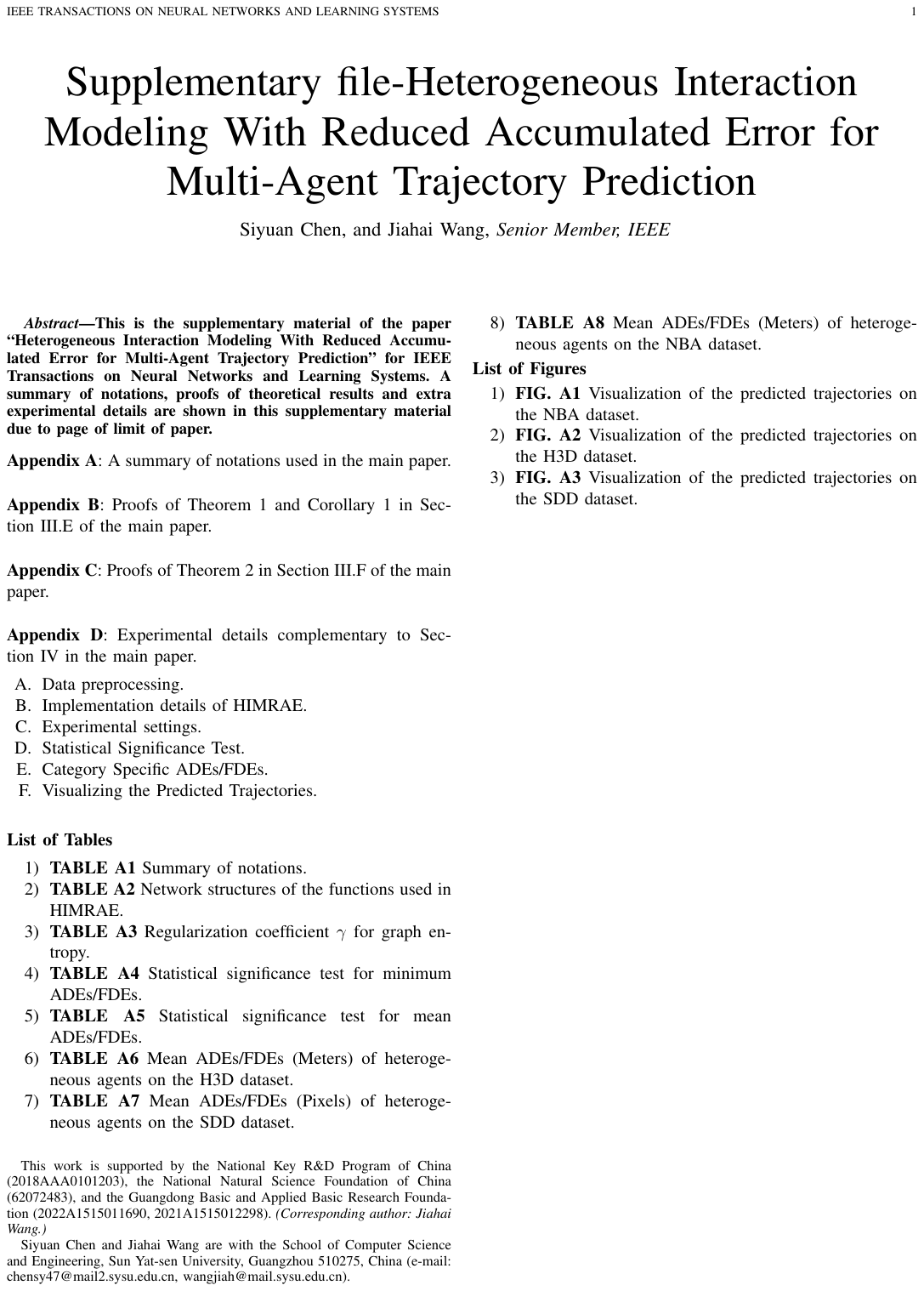}

\end{document}


\title{Supplementary file-Heterogeneous Interaction Modeling With Reduced Accumulated Error for Multi-Agent Trajectory Prediction}
\author{Siyuan Chen, and Jiahai Wang,~\IEEEmembership{Senior Member,~IEEE}
\thanks{This work is supported by the National Key R\&D Program of China (2018AAA0101203), the National Natural Science Foundation of China (62072483), and the Guangdong Basic and Applied Basic Research Foundation (2022A1515011690, 2021A1515012298). \textit{(Corresponding author: Jiahai Wang.)} }
\thanks{Siyuan Chen and Jiahai Wang are with the School of Computer Science and Engineering, Sun Yat-sen University, Guangzhou 510275, China (e-mail: chensy47@mail2.sysu.edu.cn, wangjiah@mail.sysu.edu.cn).}
}

\markboth{IEEE Transactions on Neural Networks and Learning Systems}%
{Shell \MakeLowercase{\textit{et al.}}: A Sample Article Using IEEEtran.cls for IEEE Journals}


\maketitle

\begin{abstract}
  This is the supplementary material of the paper ``Heterogeneous Interaction Modeling With Reduced Accumulated Error for Multi-Agent Trajectory Prediction'' for IEEE Transactions on Neural Networks and Learning Systems. A summary of notations, proofs of theoretical results and extra experimental details are shown in this supplementary material due to page of limit of paper.
\end{abstract}

\noindent\textbf{Appendix~A}: A summary of notations used in the main paper.

~\\
\noindent\textbf{Appendix~B}: Proofs of Theorem~1 and Corollary~1 in Section~III.E of the main paper.

~\\
\noindent\textbf{Appendix C}: Proofs of Theorem~2 in Section~III.F of the main paper.

~\\
\noindent\textbf{Appendix D}: Experimental details complementary to Section~IV in the main paper.
\begin{itemize}
  \item [A.] Data preprocessing.
  \item [B.] Implementation details of HIMRAE.
  \item [C.] Experimental settings.
  \item [D.] Statistical Significance Test.
  \item [E.] Category Specific ADEs/FDEs.
  \item [F.] Visualizing the Predicted Trajectories.
\end{itemize}

~\\
\noindent\textbf{List of Tables}

\begin{enumerate}
  \item \textbf{TABLE A1} Summary of notations.
  \item \textbf{TABLE A2} Network structures of the functions used in HIMRAE.
  \item \textbf{TABLE A3} Regularization coefficient $\gamma$ for graph entropy.
  \item \textbf{TABLE A4} Statistical significance test for minimum ADEs/FDEs.
  \item \textbf{TABLE A5} Statistical significance test for mean ADEs/FDEs.
  \item \textbf{TABLE A6} Mean ADEs/FDEs (Meters) of heterogeneous agents on the H3D dataset.
  \item \textbf{TABLE A7} Mean ADEs/FDEs (Pixels) of heterogeneous agents on the SDD dataset.
  \item \textbf{TABLE A8} Mean ADEs/FDEs (Meters) of heterogeneous agents on the NBA dataset.
\end{enumerate}

\noindent\textbf{List of Figures}
\begin{enumerate}
  \item \textbf{FIG. A1} Visualization of the predicted trajectories on the NBA dataset.
  \item \textbf{FIG. A2} Visualization of the predicted trajectories on the H3D dataset.
  \item \textbf{FIG. A3} Visualization of the predicted trajectories on the SDD dataset.
\end{enumerate}

\clearpage
\appendices
\renewcommand\thefigure{A\arabic{figure}}
\renewcommand\thetable{A\arabic{table}}
\renewcommand\theequation{A\arabic{equation}}
\setcounter{page}{1}
\setcounter{figure}{0}
\setcounter{table}{0}
\setcounter{equation}{0}
\setcounter{footnote}{0}
\setcounter{theorem}{0}
\setcounter{proposition}{0}
\setcounter{corollary}{0}

\section{Notations}\label{sec:app_notations}
Notations used in the main paper are summarized in Table~\ref{tab:notations}.
\begin{table*}[b]
  \renewcommand{\arraystretch}{1.3}
    \centering
    \caption{Summary of Notations.}
      \begin{tabularx}{\textwidth}{l|X}
      \hline
      \multicolumn{1}{l|}{Notations} & \multicolumn{1}{c}{Descriptions} \\
      \hline
      $\mathbf{x}^t_i=(x^t_i,y^t_i)$ & The 2D coordinate of agent $i$ at time $t$.\\
      $c_i$ & The category of agent $i$. \\
      $T_h,T_f$ & Historical steps and future steps. \\
      $\mathbf{X}^{1:T_h}=\{\mathbf{x}^{1:T_h}_i\}^N_{i=1}$ & The trajectories of $N$ agents from time $t=1$ to $t=T_h$. \\
      $\widehat{\mathbf{X}}^t$ & Predicted positions at time $t$.\\
      $\mathcal{G}^t=(\mathcal{V},\mathcal{E}^t)$ & A directed interaction graph at time $t$, with each agent $i\in\mathcal{V}$ as a node and the interacting relation from agent $i$ to agent $j$ as an directed edge $(i,j)\in\mathcal{E}^t$. \\
      $\tau$ & The size for a time window.\\
      $M=\lfloor (T_h+T_f)/\tau \rfloor$ & The maximum number of the time window. \\
      \hline
      $\mathbf{Z}^m\in\{0,1\}^{N\times N}$ & The interacting relations at time $t$, with $z^m_{ij}=1$ indicating a directed edge $(i, j)$. \\
      $\mathbf{E}^m\in\mathbb{R}^{N\times N\times D}$ & The interacting effects at time $t$, with $\mathbf{e}^m_{ij}$ representing the impact of agent $i$ on agent $j$.\\
      $q_\phi\left(\mathbf{Z}^m,\mathbf{E}^m|\mathbf{X}^{1:m\tau}\right)$ & The encoder parameterized by $\phi$. \\
      $p_\theta\left(\mathbf{X}^{1+m\tau:(m+1)\tau}|\mathbf{X}^{1:m\tau},\mathbf{Z}^m,\mathbf{E}^m\right)$ & The decoder parameterized by $\theta$.\\
      \hline
      $f_{\text{dec}}$ & A simplified notation of the decoder.\\
      $f_{\text{emb}}$ & An MLP for trajectory embedding.\\
      $f_v$ & An MLP updating the node embeddings.\\
      $f_e, \widetilde{f}_e,g_e$ & MLPs updating the edge embeddings.\\
      $f_{\text{proj}}$ & An MLP projecting the edge representations to a 1D space.\\
      $f_{V}$ & An MLP updating the value vector.\\
      $f_{Q}, f_{K}$ & Single-layer perceptrons updating the query vector and the key vector respectively.\\
      $g^{c_i}_{Q}, g^{c_i}_{K}, g^{c_i}_{V}$ & Category-aware single-layer perceptrons in the heterogeneous attention mechanism.\\
      $\text{GRU}_{c_j}$ & A category-aware gated recurrent unit. \\
      $f_{\text{out}}$ & An MLP predicting the change in positions. \\
      \hline
      $\mathbf{v}^m_j$ & The trajectory embedding of agent $j$ in the $m$-th time window.\\
      $\widetilde{\mathbf{v}}^m_j,\widetilde{\mathbf{e}}^m_{ij}$ & Intermediate node representations and edge representations in the encoder.\\
      $\mathbf{r}^m_{ij}$ & The hidden state for the interacting relations of edge $(i,j)$.\\
      $\mathbf{m}^t_j$ & The aggregated effects from the neighbors of agent $j$ at time $t$.\\
      $\mathbf{h}^t_j$ & The hidden state of agent $j$ at time $t$ in the decoder.\\
      \hline
      $\text{H}_{\mathcal{G}}[\mathbf{Z}^m]$ & The graph entropy. \\
      $\gamma$ & The regularization coefficient for the graph entropy. \\
      $\text{Mix}_\lambda(\cdot, \cdot)$ & The mixup operator, with $\lambda$ as the mixing coefficient. \\
      $\bar{\mathbf{X}}^t$ & A convex combination of the predicted position $\widehat{\mathbf{X}}^t$ and the ground truth $\mathbf{X}^t$. \\
      \hline
      \end{tabularx}%
    \label{tab:notations}%
  \end{table*}%

\section{Theoretical Analysis for the Graph Entropy}\label{sec:app_graph_entropy_minimizer}
  This section analyzes the effect of the graph entropy from the optimization perspective. There are two main results: (1) For a graph with a given number of edges, the graph entropy is minimized when the edges are centered on a few nodes. (2) A graph with fewer edges has a smaller lower bound of graph entropy. The proof relies on a useful inequality from the majorization theory~\cite{marshall1979inequalities}.

  Recall that the graph entropy used in this paper is defined as
  \begin{equation}
    \text{H}_{\mathcal{G}}[\mathbf{Z}^m]= -\frac{1}{\ln N}{\textstyle\sum^{N}_{j=1}(d^m_{j}/|\mathcal{E}^m|)\ln (d^m_{j}/|\mathcal{E}^m|),}
    \label{eq:duplicate_graph_entropy}
  \end{equation}
  where $\mathbf{Z}^m$ is the adjacency matrix of $N$ nodes in the $m$-th time window, $d^m_j$ is the in-degree of node $j$. For simplicity, this paper omits the superscript $m$ and the normalization constant $\frac{1}{\ln N}$ that does not affect the optimization in the following text.

  Let $\mathbf{p}=\frac{1}{|\mathcal{E}|}\left(d_1,\dots,d_N\right)$, the graph entropy in Eq.~(\ref{eq:duplicate_graph_entropy}) can be rewritten as,
  \begin{equation}
    \text{H}_{\mathcal{G}}[\mathbf{Z}]= - \sum^N_{i=1} p_i \ln p_i,
    \label{eq:simple_graph_entropy}
  \end{equation}
where $p_i$ is the $i$-th element of $\mathbf{p}$.
  
  The graph entropy defined in Eq.~(\ref{eq:simple_graph_entropy}) is essentially the Shannon entropy. It is well-known that Shannon entropy is maximized for the uniform distribution and minimized for the single-point distribution. Therefore, it can be conjectured that a graph with less spread-out edges has a lower entropy. This paper formalizes the intuition via the majorization theory.
  
  For any vector $\mathbf{x}\in \mathbb{R}^N$, let $\mathbf{x}^{\downarrow}\triangleq(x_{[1]},\dots,x_{[N]})$ denote its decreasing order, i.e., $x_{[1]}\geq\cdots\geq x_{[N]}$. Then, the definition of majorization and a useful inequality are introduced are follows.
  
  \begin{definition}[Majorization]
     For any vectors $\mathbf{x},\mathbf{y}\in\mathbb{R}^N$, we say $\mathbf{x}$ majorizes $\mathbf{y}$ (or $\mathbf{y}$ is majorized by $\mathbf{x}$), denoted by $\mathbf{x}\succ \mathbf{y}$ (or $\mathbf{y}\prec \mathbf{x}$), if
     \begin{align*}
       \sum^k_{i=1} x_{[i]} &\geq \sum^k_{i=1} y_{[i]}, 1\leq k < N,\\
       \sum^N_{i=1} x_{[i]} &= \sum^N_{i=1} y_{[i]}.
     \end{align*}
  \end{definition}
  \begin{lemma}[Hardy-Littlewood-P\'olya inequality~\cite{marshall1979inequalities}] Let $\phi$ be a concave function over $\Omega\subseteq\mathbb{R}^N$. For any vectors $\mathbf{x}, \mathbf{y}\in\Omega$, if $\mathbf{x}\succ \mathbf{y}$, then
    \begin{equation*}
      \sum^N_{i=1} \phi(x_i) \leq \sum^N_{i=1} \phi(y_i).
    \end{equation*}
    \label{lemma:majorization}
  \end{lemma}
  In our case, $\Omega=\{\mathbf{x}\in\mathbb{R}^N_{\geq 0}:\mathbf{1}^T\mathbf{x}=1\}$ is a probability simplex and $\phi(x)=-x\ln x$ is the entropy function with $\phi(0)=0$. Then, the problem of finding the minimizer of the graph entropy reduces to finding the minimum elements of $\Omega$ if they exist. Denote $\Omega(N,|\mathcal{E}|)=\left\{\frac{1}{|\mathcal{E}|}\left(d_1,\dots,d_N\right)\in\mathbb{R}^N_{\geq 0}:|\mathcal{E}|=\sum^N_{i=1}d_i\right\}$  as the probability simplex for a graph with $N$ nodes and $|\mathcal{E}|$ edges. The minimizer and the maximizer of the graph entropy are characterized as follows.
  
  \begin{theorem}[Minimizer of Graph Entropy]
    For an $N$-node directed graph with edge counts $|\mathcal{E}|$, $\min \textnormal{H}_{\mathcal{G}}[\mathbf{Z}]=\frac{k(N-1)}{|\mathcal{E}|}\ln \frac{|\mathcal{E}|}{N-1}-\frac{e}{|\mathcal{E}|}\ln \frac{e}{|\mathcal{E}|}$, where $k$ and $e$ are determined by $|\mathcal{E}|=k(N-1)+e,0\leq e < N-1$. The minimum is achieved when
    \begin{equation}
      \mathbf{p}^{\downarrow}=\Big(\underbrace{\frac{N-1}{|\mathcal{E}|},\dots,\frac{N-1}{|\mathcal{E}|}}_{k},\frac{e}{|\mathcal{E}|},\underbrace{0,\dots,0}_{N-k-1}\Big).
      \label{eq:min_graph_entropy}
    \end{equation}
    Specifically, $\min \textnormal{H}_{\mathcal{G}}[\mathbf{Z}]=0$ for $|\mathcal{E}|\leq N-1$. The minimum is achieved when $\mathbf{p}^{\downarrow}=(1,0,\dots,0)$.
  \label{thm:min_graph_entropy}
  \end{theorem}
  
  Before the proof of Theorem~\ref{thm:min_graph_entropy}, this paper first introduces a useful lemma that will be used repeatedly.
  
  \begin{lemma}
    For any vectors $\mathbf{x},\mathbf{y}\in\Omega_c=\{\mathbf{x}\in\mathbb{R}^N_{\geq 0}:\mathbf{1}^T\mathbf{x}=c\}$ with $c$ as a constant, if there exists an integer $\ell$ such that $x_{[i]}\geq y_{[i]},1\leq i\leq \ell$ and $x_{[i]}\leq y_{[i]},\ell+1\leq i\leq N$, then $\mathbf{x}\succ\mathbf{y}$.
    \label{lemma:sufficient_majorization}
  \end{lemma}
  \begin{proof}
    Since $x_{[i]}\geq y_{[i]},1\leq i\leq \ell$
    \begin{equation*}
      \sum^j_{i=1} x_{[i]} \geq \sum^j_{i=1} y_{[i]}, 1\leq j \leq \ell.
    \end{equation*}
    On the other hand, $x_{[i]}\leq y_{[i]},\ell+1\leq i\leq N$. Thus,
    \begin{equation*}
      \sum^N_{i=j} x_{[i]} \leq \sum^N_{i=j} y_{[i]}, \ell< j \leq N,
    \end{equation*}
    which implies that
    \begin{equation*}
      \sum^j_{i=1} x_{[i]} 
      = c - \sum^N_{i=j+1} x_{[i]}\\
      \geq c - \sum^N_{i=j+1} y_{[i]}\\
      =\sum^j_{i=1} y_{[i]}, \ell< j \leq N.
    \end{equation*}
    Hence, we prove that $\mathbf{x}\succ\mathbf{y}$.
  \end{proof}
  
  The constant $c$ in Lemma~\ref{lemma:sufficient_majorization} is equal to $1$ for probability simplex. The proof of Theorem~\ref{thm:min_graph_entropy} is presented as follows.
  \begin{proof}
  For any probability vector $\mathbf{q}\in\Omega(N,|\mathcal{E}|)$, this paper will prove that $\mathbf{p}\succ\mathbf{q}$. Firstly, note that $p_{[i]}\geq q_{[i]}$ for $1\leq i\leq k$ and $p_{[i]}\leq q_{[i]}$ for $k< i\leq N$. Set
  \begin{equation*}
    \ell=\begin{cases}
      k\quad&\text{if}\,\,p_{[k+1]}\leq q_{[k+1]},\\
      k+1\quad&\text{otherwise.}
    \end{cases}
  \end{equation*}
  Then $\ell$ satisfies the condition in Lemma~\ref{lemma:sufficient_majorization}, which implies that  $\mathbf{p}\succ\mathbf{q}$. It can be evaluated that
  \begin{align*}
    &\quad\,\min \textnormal{H}_{\mathcal{G}}[\mathbf{Z}]\\ &= -\sum^k_{i=1}\frac{N-1}{|\mathcal{E}|}\ln \frac{N-1}{|\mathcal{E}|}-\frac{e}{|\mathcal{E}|}\ln \frac{e}{|\mathcal{E}|}-\sum^{N-k-1}_{i=1}\phi(0)\\
    &= \frac{k(N-1)}{|\mathcal{E}|}\ln \frac{|\mathcal{E}|}{N-1}-\frac{e}{|\mathcal{E}|}\ln \frac{e}{|\mathcal{E}|}.
  \end{align*}
  When $|\mathcal{E}|\leq N-1$, one finds $k=0$ and $e=|\mathcal{E}|$. In this case, $\mathbf{p}^{\downarrow}=(1,0,\dots,0)$ and $\min \textnormal{H}_{\mathcal{G}}[\mathbf{Z}]=0$.
  \end{proof}
  
  \begin{corollary}
    Given the number of nodes, a graph with fewer edges has a smaller lower bound of graph entropy.
  \end{corollary}
  
  \begin{proof}
    Let $\mathbf{p}$ and $\mathbf{q}$ be the minimum elements of $\Omega(N,|\mathcal{E}|)$ and $\Omega(N,|\mathcal{E}'|)$, respectively. Suppose that $|\mathcal{E}|=k(N-1)+e\leq |\mathcal{E}'|=k'(N-1)+e'$. One can verify that $k'\geq k$, $p_{[i]}\geq q_{[i]},1\leq i\leq k$ and $p_{[i]}\leq q_{[i]},k+2\leq i\leq N$. Set
    \begin{equation*}
      \ell=\begin{cases}
        k\quad&\text{if}\,\,p_{[k+1]}\leq q_{[k+1]},\\
        k+1\quad&\text{otherwise.}
      \end{cases}
    \end{equation*}
    Then $\ell$ satisfies the condition in Lemma~\ref{lemma:sufficient_majorization}, which implies that  $\mathbf{p}\succ\mathbf{q}$. Hence $\min\textnormal{H}_{\mathcal{G}}[\mathbf{Z}]\leq \min\textnormal{H}_{\mathcal{G}}[\mathbf{Z}']$.
  \end{proof}

\section{Proof of Theorem~2}\label{sec:app_proof}
\begin{theorem}
    Let $f$ be a learned model with bounded single-step prediction error $\|f(\mathbf{X}^t)-\mathbf{X}^{t+1}\|\leq \epsilon$. Assume that $f$ is Lipschitz continuous with constant $L_f$ under the norm $\|\cdot\|$. Then, the following upper bounds hold,
    \begin{itemize}
        \item[(1)] $\|f^n(\widehat{\mathbf{X}}^t)-\mathbf{X}^{t+n}\|\leq L_f^n\|\widehat{\mathbf{X}}^t-\mathbf{X}^t\|+\frac{L_f^n-1}{L_f-1}\epsilon$;
        \item[(2)] $\underset{\lambda}{\mathbb{E}}[\|f^n(\bar{\mathbf{X}}^t)-\mathbf{X}^{t+n}\|]\leq \frac{1}{2}L_f^n\|\widehat{\mathbf{X}}^t-\mathbf{X}^t\|+\frac{L_f^n-1}{L_f-1}\epsilon$;
        \item[(3)] $\underset{\lambda}{\mathbb{E}}[\|f^n(\widehat{\mathbf{X}}^t)-f^n(\bar{\mathbf{X}}^t)\|]\leq \frac{1}{2}L_f^n\|\widehat{\mathbf{X}}^t-\mathbf{X}^t\|$.
    \end{itemize}
\end{theorem}
\begin{proof}
Proposition (1) can be proved by induction. Firstly,
\begin{equation*}
\begin{aligned}
    &\quad\,\,\|f^n(\widehat{\mathbf{X}}^t)-\mathbf{X}^{t+n}\|\\
    &= \|f^n(\widehat{\mathbf{X}}^t)-f(\mathbf{X}^{t+n-1})+f(\mathbf{X}^{t+n-1})-\mathbf{X}^{t+n}\|\\
    &\leq \|f^n(\widehat{\mathbf{X}}^t)-f(\mathbf{X}^{t+n-1})\|+\|f(\mathbf{X}^{t+n-1})-\mathbf{X}^{t+n}\|\\
    &\leq L_f\|f^{n-1}(\widehat{\mathbf{X}}^t)-\mathbf{X}^{t+n-1}\|+\epsilon
\end{aligned}
\end{equation*}
where the first inequality is due to the triangle inequality, and the second inequality is due to the assumptions in the theorem. Denoting $\epsilon_{n}=\|f^n(\widehat{\mathbf{X}}^t)-\mathbf{X}^{t+n}\|$, the result above can be rewritten as
\begin{equation*}
\epsilon_{n}\leq L_f\epsilon_{n-1}+\epsilon,n\geq 1.
\end{equation*}
By induction, it can be proved that
\begin{equation*}
\|f^n(\widehat{\mathbf{X}}^t)-\mathbf{X}^{t+n}\|\leq L_f^n\|\widehat{\mathbf{X}}^t-\mathbf{X}^t\|+\frac{L_f^n-1}{L_f-1}\epsilon.
\end{equation*}
This completes the proof of proposition (1). A similar result can be obtained for proposition (2) as follows,
\begin{equation*}
\begin{aligned}
    \|f^n(\widehat{\mathbf{X}}^t)-\mathbf{X}^{t+n}\|
    &\leq L_f^n\|\bar{\mathbf{X}}^t-\mathbf{X}^t\|+\frac{L_f^n-1}{L_f-1}\epsilon\\
    &= \lambda L_f^n\|\widehat{\mathbf{X}}^t-\mathbf{X}^t\|+\frac{L_f^n-1}{L_f-1}\epsilon,
\end{aligned}
\end{equation*}
where the equality is by the definition of $\bar{\mathbf{X}}^t$. Note that $\underset{\lambda}{\mathbb{E}}[\lambda]=\frac{1}{2}$ for $\lambda\sim\text{Beta}(\alpha,\alpha)$. Taking expectation w.r.t. $\lambda$ from both sides, one has
\begin{equation*}
    \underset{\lambda}{\mathbb{E}}[\|f^n(\widehat{\mathbf{X}}^t)-\mathbf{X}^{t+n}\|]
    \leq \frac{1}{2} L_f^n\|\widehat{\mathbf{X}}^t-\mathbf{X}^t\|+\frac{L_f^n-1}{L_f-1}\epsilon.
\end{equation*}

For proposition (3), by recursively using the triangle inequality, 
\begin{equation*}
\|f^n(\widehat{\mathbf{X}}^t)-f^n(\bar{\mathbf{X}}^t)\|\leq (1-\lambda)L_f^n\|\widehat{\mathbf{X}}^t-\mathbf{X}^t\|.
\end{equation*}
Again, by taking expectation w.r.t. $\lambda$ from both sides, one has
\begin{equation*}
\underset{\lambda}{\mathbb{E}}[\|f^n(\widehat{\mathbf{X}}^t)-f^n(\bar{\mathbf{X}}^t)\|]\leq \frac{1}{2}L_f^n\|\widehat{\mathbf{X}}^t-\mathbf{X}^t\|.
\end{equation*}
\end{proof}

\section{Experiments}\label{sec:app_experiments}
\subsection{Data Preprocessing}
This subsection describes the supplementary details of data preprocessing. The raw data of NBA\footnote{https://github.com/linouk23/NBA-Player-Movements} and SDD\footnote{https://cvgl.stanford.edu/projects/uav\_data/} can be directly downloaded as they are public online. The raw data of H3D\footnote{https://usa.honda-ri.com/h3d} along with necessary codes for object detection and tracking can be requested from the Honda Research Institute. 

The raw data are preprocessed by ourselves according to the instructions of the original paper~\cite{li2020evolvegraph}. Regarding each game in NBA as a scene, all three datasets contain multiple scenes. For each scene, the trajectories are down-sampled at the rate of 2.5Hz. All types of agents are preserved and the trajectories of all agents within a given period ($15$ for NBA and H3D datasets, and $20$ for the SDD dataset) are treated as a sample. The number of agents in a sample does not change over time, while the agent counts may vary in different samples. Non-overlapping trajectories from each scene constitute the training set, validation set and testing set with ratios 0.65, 0.10 and 0.25, respectively.

As shown in Table~I, the numerical ranges of the 2D coordinates vary significantly across different datasets. To ensure numerical stability, the coordinates are normalized to $[-1,1]$ via the min-max normalization, and they are recovered while making a prediction.

Context information like the top-down-view images and point cloud density maps are not included in the experiments. Incorporating the context information is left for future work.

\subsection{Implementation Details of HIMRAE}\label{sec:app_implementations}
Most functions used in HIMRAE are single-layer perceptrons or MLPs. The network structure of these functions are listed in Table~\ref{tab:network_structure}. Other implementation details of HIMRAE are listed as follows.
\begin{itemize}
  \item Following Li~\textit{et al.}~\cite{li2020evolvegraph}, two-layer GRUs are used in both the encoder and the decoder.
  \item The temperature parameter $T$ in Eq.~(8) is set to $0.5$.
  \item The variance $\sigma^2$ is set to $1$ for the distribution of $\widehat{\mathbf{X}}^t$.
\end{itemize}
\begin{table}[t]
  \renewcommand{\arraystretch}{1.3}
  \centering
  \caption{Network structures of the functions used in HIMRAE.}
    \begin{tabularx}{\columnwidth}{c|c}
    \hline
    Function & Network Structure  \\
    \hline
    $f_{\text{emb}},f_v,f_e,\widetilde{f}_e$ & [Linear, ELU, BatchNorm1d]$\times 2$  \\
    \hline
    $f_{Q},f_{K},g^{c_i}_{Q}, g^{c_i}_{K}, g^{c_i}_{V}$ & [Linear, Tanh]  \\
    \hline
    $f_{V}$ & [Linear, Tanh]$\times 2$ \\
    \hline
    $f_{\text{proj}}$ & [[Linear, ELU, BatchNorm1d]$\times 2$, Linear] \\
    \hline
                  $f_{\text{out}}$ & [[Linear, ReLU]$\times 2$, Linear] \\
    \hline
    \end{tabularx}
  \label{tab:network_structure}%
\end{table}
\subsection{Experimental Settings}
\begin{table}[t]
  \renewcommand{\arraystretch}{1.3}
  \centering
  \caption{Regularization coefficient $\gamma$ for graph entropy.}
    \begin{tabular}{c|c|c|c}
    \hline
    Dataset & Candidate Values for $\gamma$ & Best $\gamma$ & Best $\gamma$ w/mixup   \\
    \hline
    NBA & $\{10^{-6}, 10^{-5}, 10^{-4}, 10^{-3}\}$ & $10^{-4}$ & $10^{-5}$ \\
    \hline
    H3D & $\{10^{-5}, 10^{-4}, 10^{-3}, 10^{-2}\}$ & $10^{-3}$ & $10^{-4}$ \\
    \hline
    SDD & $\{0.1, 1, 10, 10^2\}$ & $10$ & $10$ \\
    \hline
    \end{tabular}
  \label{tab:gamma_candidate}%
\end{table}
\begin{itemize}
    \item All experiments are run 5 times on a machine containing 128GB of RAM, and 8 NVIDIA TITANXP graphics cards with PyTorch 1.9 and CUDA 11.5 in Ubuntu 20.04.
    \item Graphs of different sizes are batched into a single graph with different connected components, which allows better usage of the memory and significantly reduces the training time.
    \item The dimensions of all hidden layers are set to $128$. 
    \item Our model is trained with a batch size of 128 for 200 epochs. The Adam optimizer is adopted with a learning rate of $10^{-3}$.
    \item Empirically, the windows size $\tau$ is set to $5$ for NBA and H3D datasets, and it is set to $4$ for the SDD dataset. Experiments by Li~\textit{et al.}~\cite{li2020evolvegraph} showed that, a smaller $\tau$ leads to lower predicting errors, while more inference time is required. Therefore, an intermediate value can achieve a good trade-off.
    \item For the graph entropy, the regularization coefficient $\gamma$ is chosen via cross-validation. The candidate set, the best $\gamma$, and the best $\gamma$ used in HIMRAE w/GE \& mixup are listed in Table~\ref{tab:gamma_candidate}.
    \item For the mixup training, the parameter $\alpha$ is empirically initialized as $10$ and decays by $0.5$ in every $10$ epochs. 
    $\alpha$ controls the variance of the distribution $\text{Beta}(\alpha, \alpha)$ for the mixing coefficient $\lambda$. When $\alpha$ is large, $\lambda\approx 1/2$, leading to an approximately equal-weighted mixture. In this case, the accumulated errors are more controllable. When $\alpha$ is small, more choices are allowed for $\lambda$, and the errors are less controllable, which increases the training difficulty. By starting with a relatively large $\alpha$ and decreasing it linearly, the model progressively learns to reduce the accumulated errors. Other choices of $\alpha'$s initial value and more designs for the decaying strategy may be explored to further improve the performance.
\end{itemize}
\subsection{Statistical Significance Test}
\begin{table*}[t]\renewcommand{\arraystretch}{1.3}
  \centering
  \caption{Statistical significance test for minimum ADEs/FDEs.}
    \begin{tabular}{c|c|c|c|c|c|c}
    \hline
    \multicolumn{1}{c|}{Datasets} & \multicolumn{2}{c|}{NBA} & \multicolumn{2}{c|}{H3D} & \multicolumn{2}{c}{SDD} \\
    \hline
    Units & \multicolumn{2}{c|}{Meters} & \multicolumn{2}{c|}{Meters} & \multicolumn{2}{c}{Pixels}\\
    \hline
    \multicolumn{1}{c|}{Metrics} & \multicolumn{1}{c|}{minADE} & \multicolumn{1}{c|}{minFDE} & \multicolumn{1}{c|}{minADE} & \multicolumn{1}{c|}{minFDE} & \multicolumn{1}{c|}{minADE} & \multicolumn{1}{c}{minFDE} \\
    \hline
    EvolveGraph &0.12 	&0.21  	&0.58 	&1.00  	&20.9 	&39.8 \\
    HIMRAE w/GE \& mixup &\textbf{0.09} 	&\textbf{0.17} 	 		&\textbf{0.35} 	&\textbf{0.50} 	 		&\textbf{18.8} 	&\textbf{36.4}    \\
    $p$-values & $2.22\times 10^{-4}$      & $1.84\times 10^{-5}$      & $5.51\times 10^{-11}$      & $4.50\times10^{-10}$      & $1.09\times 10^{-5}$      & $1.23\times 10^{-4}$  \\
    \hline
    \end{tabular}%
  \label{tab:pvalues_min}%
\end{table*}%

\begin{table*}[t]\renewcommand{\arraystretch}{1.3}

  \centering
  \caption{Statistical significance test for mean ADEs/FDEs.}
    \begin{tabular}{c|c|c|c|c|c|c}
    \hline
    \multicolumn{1}{c|}{Datasets} & \multicolumn{2}{c|}{NBA} & \multicolumn{2}{c|}{H3D} & \multicolumn{2}{c}{SDD} \\
    \hline
    Units & \multicolumn{2}{c|}{Meters} & \multicolumn{2}{c|}{Meters} & \multicolumn{2}{c}{Pixels}\\
    \hline
    \multicolumn{1}{c|}{Metrics} & \multicolumn{1}{c|}{minADE} & \multicolumn{1}{c|}{minFDE} & \multicolumn{1}{c|}{minADE} & \multicolumn{1}{c|}{minFDE} & \multicolumn{1}{c|}{minADE} & \multicolumn{1}{c}{minFDE} \\
    \hline
    EvolveGraph &0.19 	&0.34 	&0.81 	&1.49 	&28.3 	&54.0  \\
    HIMRAE w/GE \& mixup &\textbf{0.15} 	&\textbf{0.28} &\textbf{0.56} 	&\textbf{0.99} &\textbf{25.2} 	&\textbf{49.4} \\
    $p$-values & $9.75\times 10^{-4}$      & $1.88\times 10^{-4}$      & $5.95\times 10^{-12}$      & $1.23\times 10^{-10}$      & $1.13\times 10^{-8}$      & $5.27\times 10^{-8}$  \\
    \hline
    \end{tabular}%
  \label{tab:pvalues_mean}%
\end{table*}%

\begin{table*}[t]
  \renewcommand{\arraystretch}{1.3}
  \centering
  \caption{Mean ADEs/FDEs (Meters) of heterogeneous agents on the H3D dataset.}
    \begin{tabular}{c|c|c|c|c|c|c|c|c}
    \hline
    Type & Car & Pedestrian & Truck & Cyclist & Other vehicle & Bus & Motorcyclist & Animal \\
    \hline
    Percentage (\%) & 50.05 & 36.62 & 8.93 & 2.36 & 1.43 & 0.32 & 0.26 & 0.03 \\
    \hline
    HIMRAE${}_{\text{HOMO}}$ &0.75/1.44 	&0.69/1.23 	&0.73/\textbf{1.33} 	&0.92/1.64 	&0.71/1.34 	&1.32/1.63 	&\textbf{1.14}/\textbf{2.17} 	&0.19/0.28 \\
    HIMRAE &\textbf{0.73}/\textbf{1.39} 	&\textbf{0.66}/\textbf{1.16} 	&\textbf{0.72}/1.34 	&\textbf{0.79}/\textbf{1.14} 	&\textbf{0.69}/\textbf{1.19} 	&\textbf{0.76}/\textbf{1.14} 	&1.32/2.68 	&\textbf{0.15}/\textbf{0.20} \\
    \hline
    \end{tabular}
  \label{tab:cat_H3D}
\end{table*}
\begin{table*}[t]\renewcommand{\arraystretch}{1.3}
      \centering
      \caption{Mean ADEs/FDEs (Pixels) of heterogeneous agents on the SDD dataset.}
        \begin{tabular}{c|c|c|c|c|c|c}
        \hline
        Type & Pedestrian & Biker & Car & Cart & Skater & Bus \\
        \hline
        Percentage (\%) & 65.87 & 17.29 & 14.85 & 0.73 & 0.66  & 0.60 \\
        \hline
        HIMRAE${}_{\text{HOMO}}$ &23.8/43.7 	&67.9/130.7 	&10.7/18.6 	&\textbf{60.1}/\textbf{112.5} 	&\textbf{69.6}/\textbf{132.8} 	&69.6/132.8 \\
        HIMRAE &\textbf{23.3}/\textbf{42.7} 	&\textbf{62.7}/\textbf{122.4} 	&\textbf{9.5}/\textbf{17.0} 	&62.9/116.5 	&77.1/143.3 	&\textbf{19.7}/\textbf{34.2} \\
        \hline
        \end{tabular}
      \label{tab:cat_SDD}%
\end{table*}
\begin{table*}[!t]\renewcommand{\arraystretch}{1.3}
  \centering
  \caption{Mean ADEs/FDEs (Meters) of heterogeneous agents on the NBA dataset.}
    \begin{tabular}{c|c|c|c}
    \hline
    Type & Ball & Home Team Player & Visiting Team Player \\
    \hline
    Percentage  & 1/11 & 5/11 & 5/11 \\
    \hline
    HIMRAE${}_{\text{HOMO}}$ &0.44/0.81 	&0.15/0.26 	&0.15/0.26 \\
    HIMRAE &\textbf{0.43}/\textbf{0.79} 	&\textbf{0.13}/\textbf{0.24} 	&\textbf{0.13}/\textbf{0.23} \\
    \hline
    \end{tabular}
  \label{tab:cat_NBA}%
\end{table*}

Statistical significance tests are conducted on all datasets by comparing HIMRAE w/GE \& mixup and EvolveGraph, the best baseline overall. The $p$-values of minimum ADEs/FDEs and mean ADEs/FDEs are reported in Table~\ref{tab:pvalues_min} and Table~\ref{tab:pvalues_mean}, respectively. $p < 0.01$ indicates that two distributions are significantly different.

\subsection{Category Specific ADEs/FDEs}
To further demonstrate the effectiveness of HAM, this paper reports the mean ADEs/FDEs of HIMRAE and HIMRAE${}_{\text{HOMO}}$ on heterogeneous agents. The results on H3D, SDD and NBA datasets are shown in Table~\ref{tab:cat_H3D}, Table~\ref{tab:cat_SDD} and Table~\ref{tab:cat_NBA}, respectively. The results show that the predicting errors for almost all types of agents decrease by using HAM. On the H3D dataset, the ADEs/FDEs of some minor classes like cyclists, buses and animals decrease more significantly, which shows that the fine-grained interactions learned by HAM can help to make better predictions for rare classes. Similar results are shown on the SDD dataset except that the predicting errors for carts and skaters are larger. On the NBA dataset, the predicting error for the ball is significantly larger than that of a player. Maybe the ball moves with larger uncertainty, and its trajectory is harder to predict.

\begin{figure*}[!t]
  \centering
  \subfloat[HIMRAE w/GE \& mixup]{\includegraphics[width=0.9\textwidth]{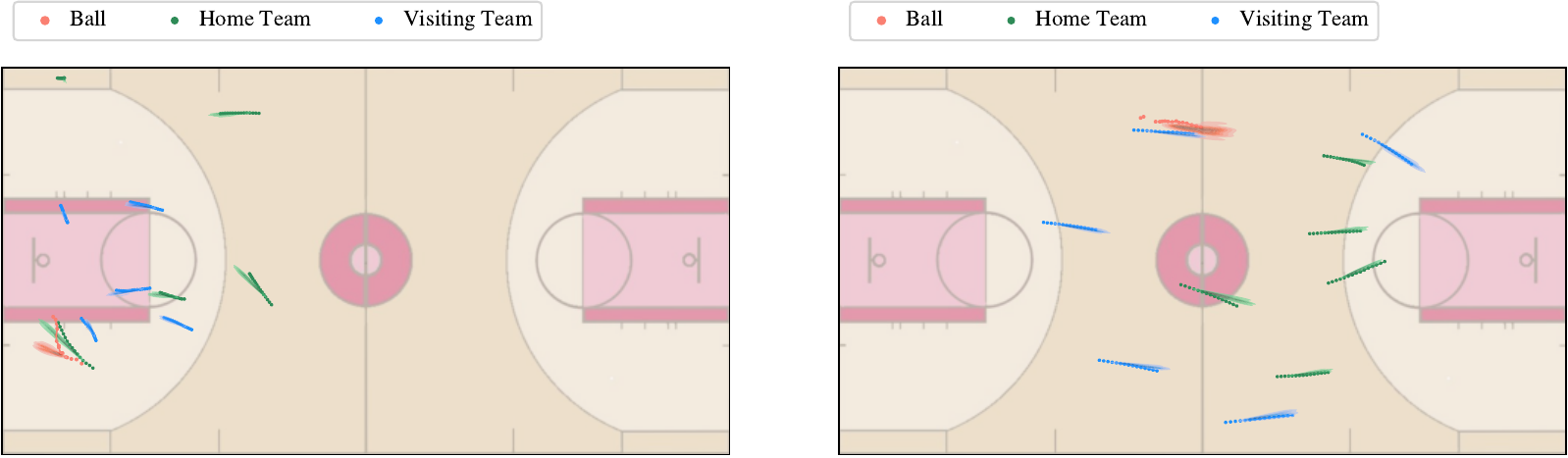}}
  \label{fig:nba_HIMRAE}
  \hfil
  \subfloat[EvolveGraph]{\includegraphics[width=0.9\textwidth]{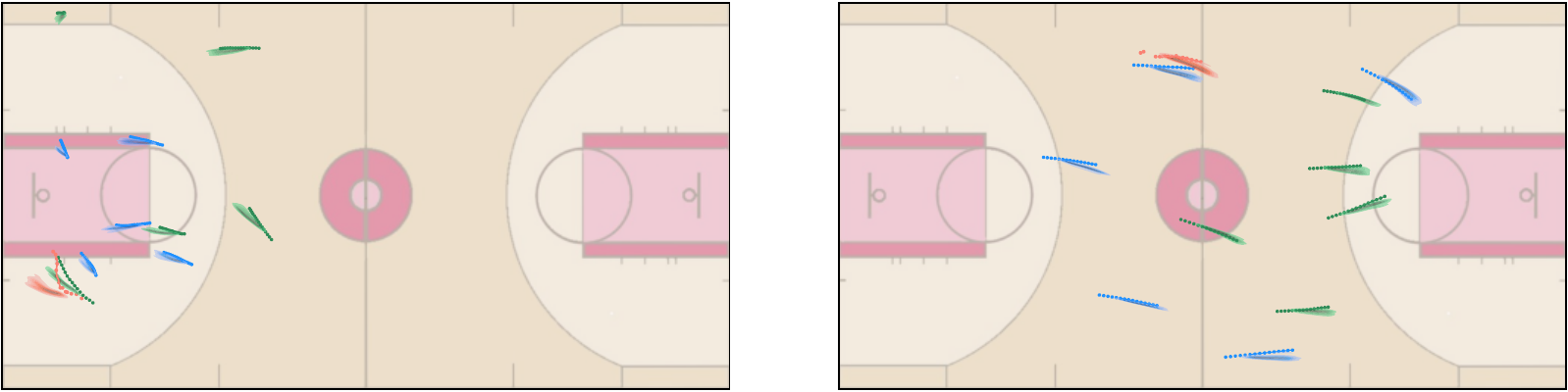}}
  \label{fig:nba_EvolveGraph}
  \caption{Visualization of the predicted trajectories on the NBA dataset. Each column corresponds to the results on a single sample. The trajectories of the ball, home team players and visiting team players are in orange, green and blue, respectively. Historical trajectories are in dots, ground truth trajectories to be predicted are in solid lines, while the predicted trajectories are visualized using the kernel density estimation.)}
  \label{fig:traj_nba}
\end{figure*}

\begin{figure*}[!t]
  \centering
  \subfloat[HIMRAE w/GE \& mixup]{\includegraphics[width=0.9\textwidth]{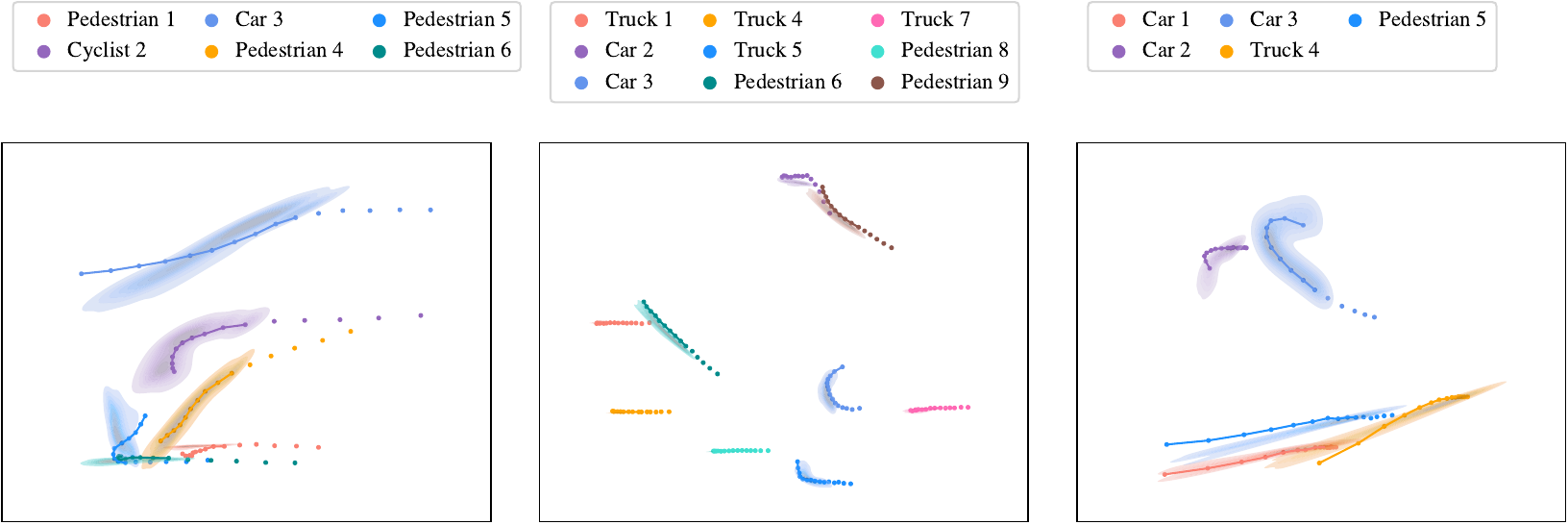}}
  \label{fig:h3d_HIMRAE}
  \hfil
  \subfloat[EvolveGraph]{\includegraphics[width=0.9\textwidth]{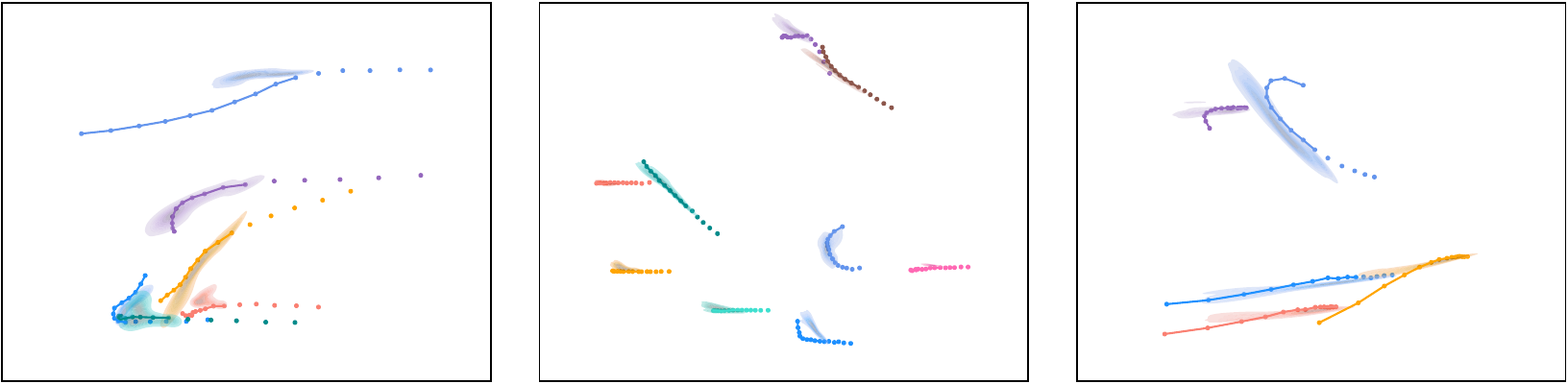}}
  \label{fig:h3d_EvolveGraph}
  \caption{Visualization of the predicted trajectories on the H3D dataset. Each column corresponds to the results on a single sample. Historical trajectories are in dots, ground truth trajectories to be predicted are in solid lines, while the predicted trajectories are visualized using the kernel density estimation.}
  \label{fig:traj_h3d}
\end{figure*}

\begin{figure*}[!t]
  \centering
  \subfloat[HIMRAE w/GE \& mixup]{\includegraphics[width=0.9\textwidth]{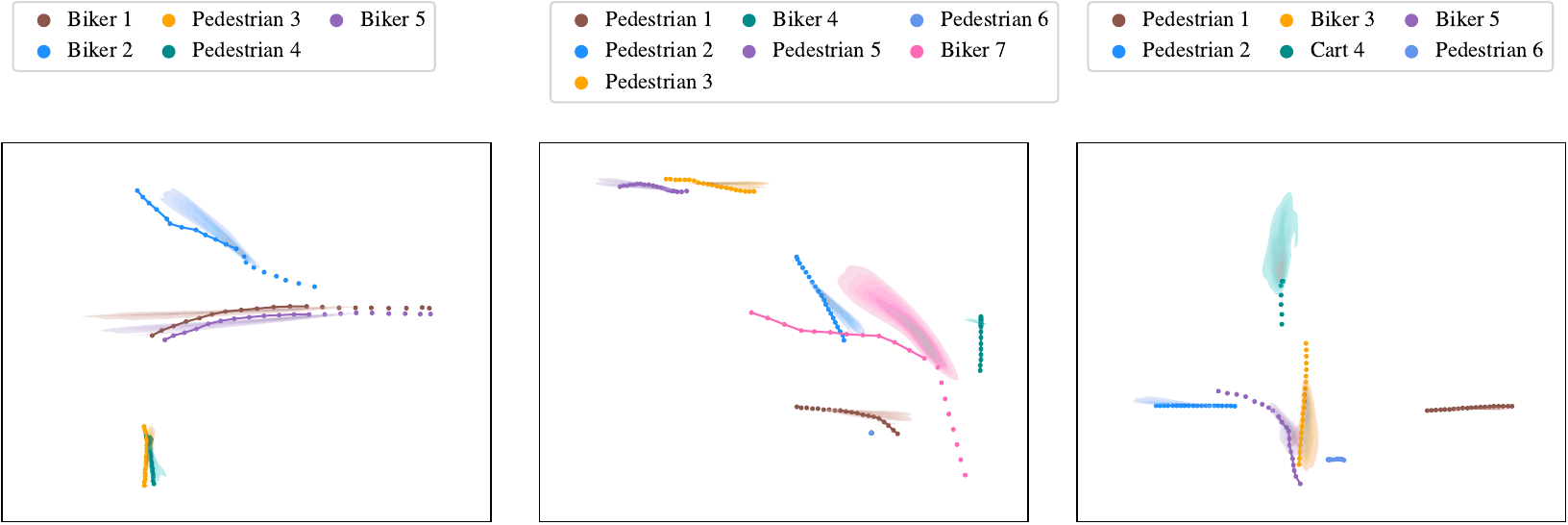}}
  \label{fig:stanford_HIMRAE}
  \hfil
  \subfloat[EvolveGraph]{\includegraphics[width=0.9\textwidth]{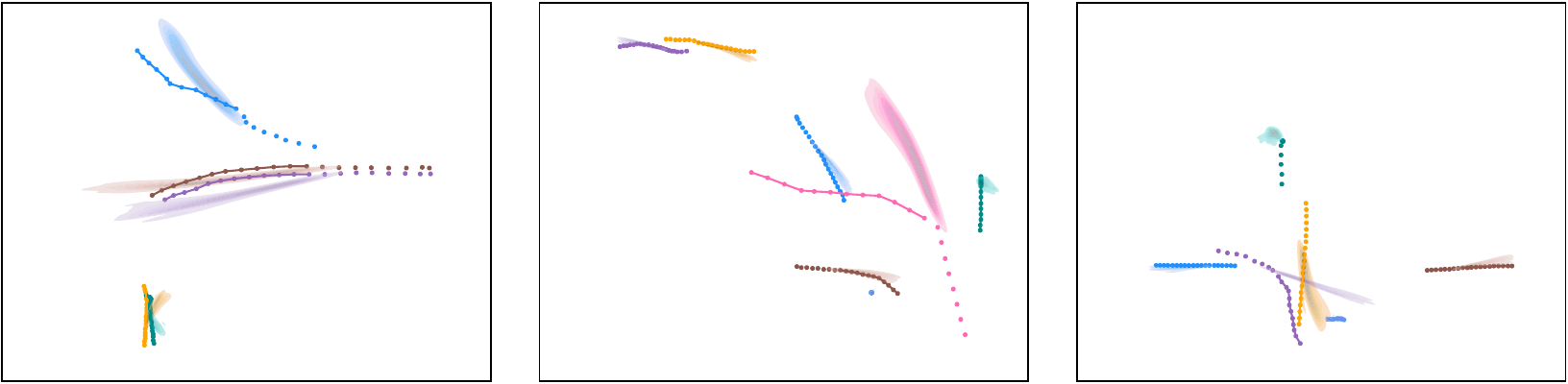}}
  \label{fig:stanford_EvolveGraph}
  \caption{Visualization of the predicted trajectories on the SDD dataset. Each column corresponds to the results on a single sample. Historical trajectories are in dots, ground truth trajectories to be predicted are in solid lines, while the predicted trajectories are visualized using the kernel density estimation.}
  \label{fig:traj_sdd}
\end{figure*}
\subsection{Visualizing the Predicted Trajectories}
This subsection visualizes the predicted trajectories of our method and EvolveGraph, a closely related work for heterogeneous multi-agent trajectory prediction. Randomly selected samples from NBA, H3D and SDD datasets are visualized in Fig.~\ref{fig:traj_nba}, Fig.~\ref{fig:traj_h3d} and Fig.~\ref{fig:traj_sdd}, respectively.

As shown in Fig.~\ref{fig:traj_nba}, our method can better predict the moving directions of both the ball and the players. In both cases, EvolveGraph fails to make accurate predictions for the ball that generally moves with large uncertainty. Consequently, the predicted trajectories of players around the ball suffer from a larger deviation than our predictions. Maybe our method can distinguish the ball-player interaction and the player-player interaction, and make more accurate predictions. 

The trajectories on the H3D dataset and the SDD dataset are more complex since more types of traffic participants are involved. Compared with EvolveGraph, the predictions of our method can better cover the target trajectories of all types of agents. Specifically, in the third case of Fig.~\ref{fig:traj_h3d}, our method successfully predicts the turning behavior of Car 3. Nonetheless, in the second case of Fig.~\ref{fig:traj_sdd}, the predictions of both methods for Biker 7 suffer from a large deviation from the ground truth. Such a case is difficult to handle since the sudden change of intention can be hardly inferred from only the historical trajectories.

\bibliographystyle{IEEEtran}
\bibliography{IEEEabrv,ref}